\documentclass[reqno,centertags, 12pt]{amsart}
\usepackage{amsmath,amsthm,amscd,amssymb}
\usepackage{latexsym}
\usepackage{graphicx}
\usepackage{subfigure}
\newcommand{\bbC}{{\mathbb{C}}}
\newcommand{\bbN}{{\mathbb{N}}}

\newcommand{\bbQ}{{\mathbb{Q}}}
\newcommand{\bbR}{{\mathbb{R}}}

\newcommand{\bbZ}{{\mathbb{Z}}}

\newcommand{\Pf}{{\rm Pf}}

\newcommand{\lb}{\label}

\newcommand{\bi}{\bibitem}

\newcommand{\beq}{\begin{equation}}
\newcommand{\eeq}{\end{equation}}
\newcommand{\ba}{\begin{align}}
\newcommand{\ea}{\end{align}}

\newcommand{\dilog}{\text{\rm{dilog}}}

\newcounter{smalllist}

\allowdisplaybreaks
\numberwithin{equation}{section}

\newtheorem{theorem}{Theorem}[section]

\newtheorem*{p2.1}{Proposition 2.1}
\newtheorem{proposition}[theorem]{Proposition}
\newtheorem{lemma}[theorem]{Lemma}
\newtheorem{corollary}[theorem]{Corollary}
\theoremstyle{definition}

\newtheorem*{tA}{Theorem A}
\newtheorem*{tB}{Theorem B}

\theoremstyle{remark}
\newtheorem*{remark}{Remark}

\theoremstyle{definition}
\newtheorem*{definition}{Definition}
\newtheorem*{claim1}{Claim 1}
\newtheorem*{claim2}{Claim 2}
\newtheorem*{claim3}{Claim 3}
\newtheorem*{claim4}{Claim 4}
\newtheorem*{claim5}{Claim 5}
\newtheorem*{propint}{Proposition}

\newtheorem*{acknowledgement}{Acknowledgment}

\begin{document}
\title[{\tiny Shifted Schur process and asymptotics of large random SPP}]{Shifted Schur process and asymptotics of large random strict plane partitions}
\author[{\tiny M. Vuleti\'{c}}]{Mirjana Vuleti\'{c}$^{*}$}

\thanks{$^*$ Mathematics 253-37, California Institute of Technology, Pasadena, CA 91125.
E-mail: vuletic@caltech.edu }

\begin{abstract}
In this paper we define the shifted Schur process as a measure on
sequences of strict partitions. This process is a generalization of
the shifted Schur  measure introduced in  \cite{TW} and \cite{Mat}
and is a shifted version of the Schur process introduced in
\cite{OR}. We prove that the shifted Schur process defines a
Pfaffian point process. We further apply this fact to compute the
bulk scaling limit of the correlation functions for a measure on
strict plane partitions which is an analog of the uniform measure on
ordinary plane partitions. As a byproduct, we obtain a shifted
analog of the famous MacMahon's formula.
\end{abstract}

\maketitle

\section{Introduction} \lb{s1}

The basic object of this paper is the {\it shifted Schur process}
that we introduce below.

Consider the following two figures:

\begin{figure}[htp!]
{\includegraphics[width=.35\textwidth]{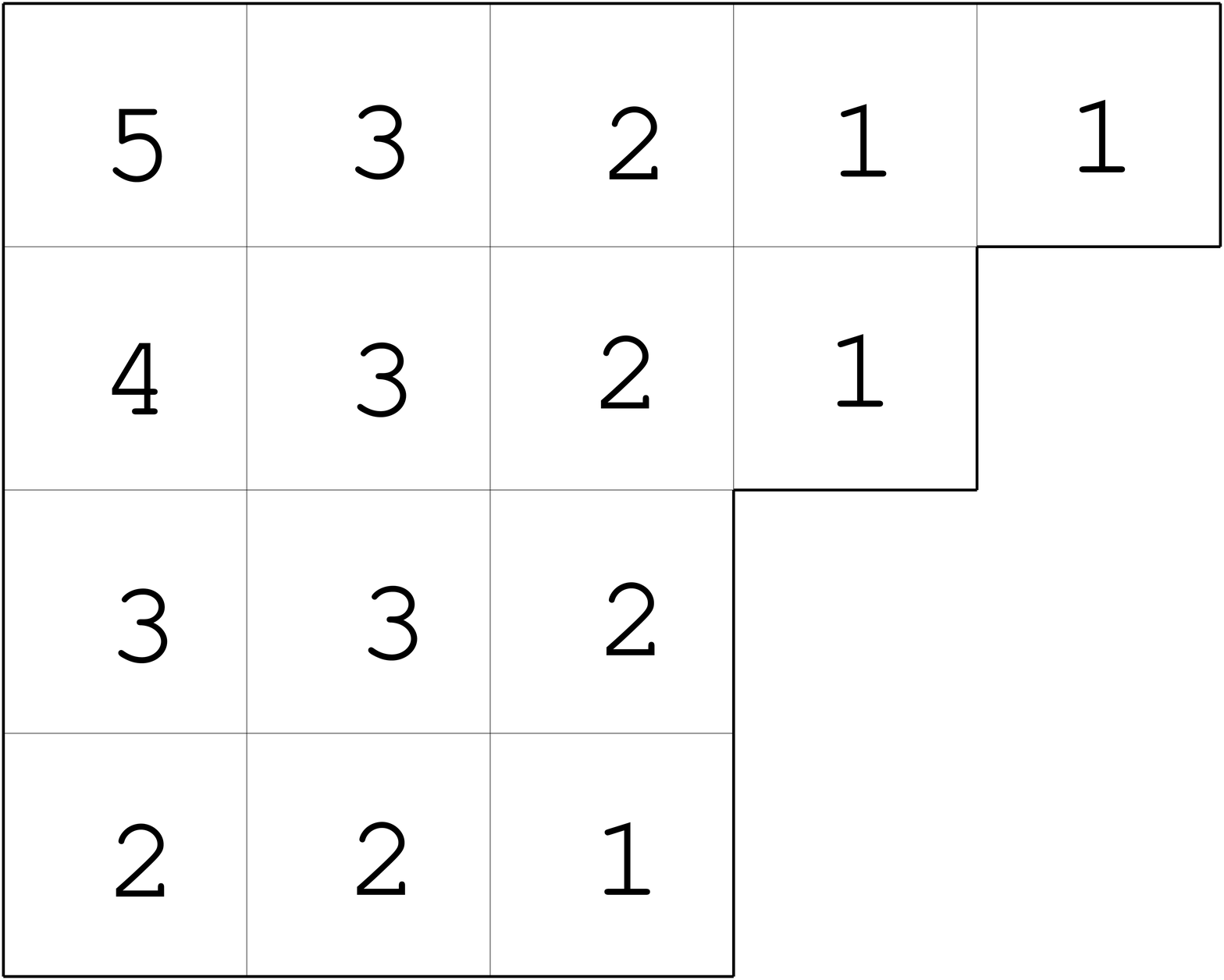}}
\hspace{1cm}
{\includegraphics[width=.35\textwidth]{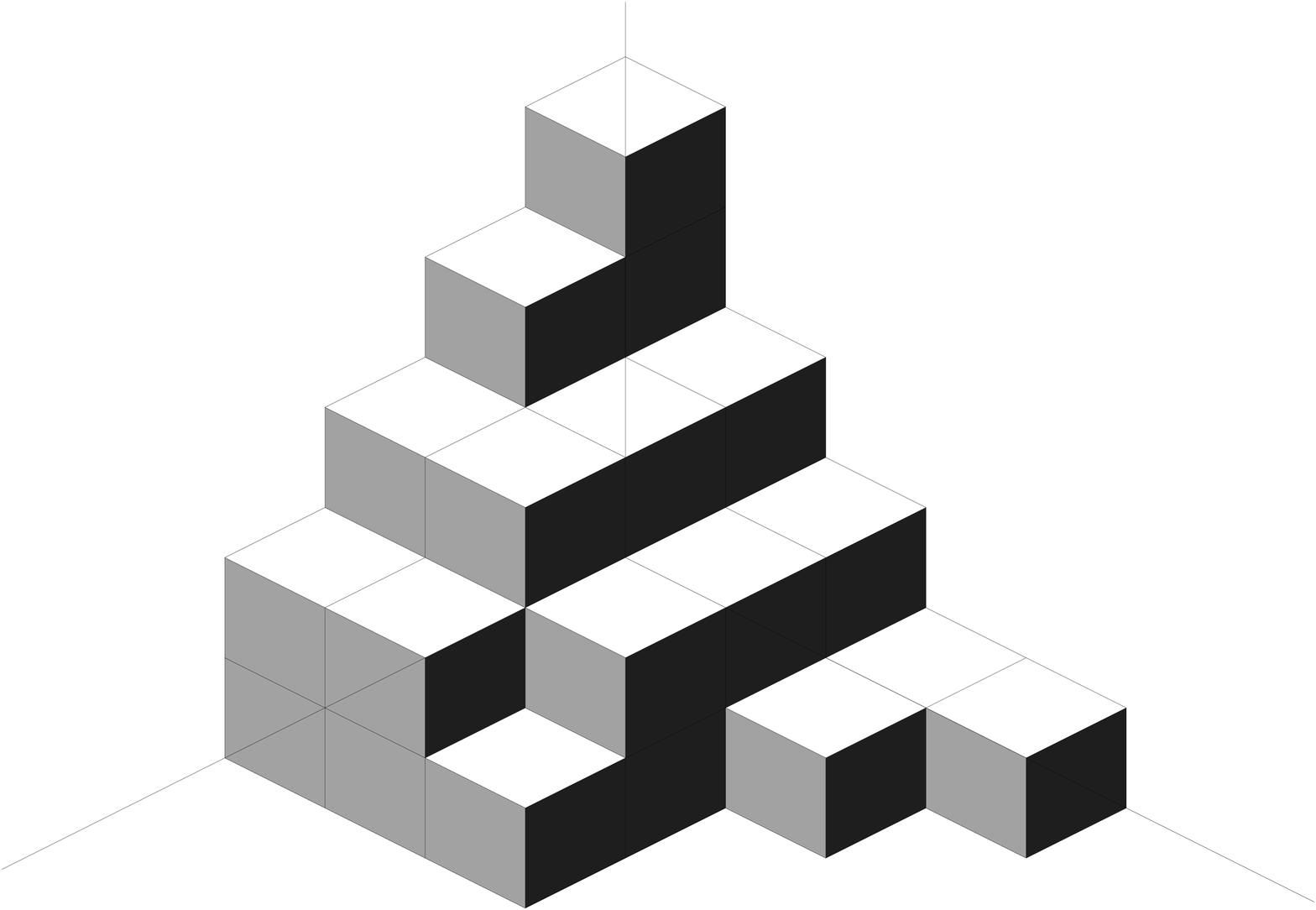}}
\end{figure}

Both these figures represent a {\it plane partition}~ -- an infinite
matrix with nonnegative integer entries that form nonincreasing rows
and columns with only finitely many nonzero entries. The second
figure shows the plane partition as a 3-dimensional object where the
height of the block positioned at $(i,j)$ is given with the
$(i,j)$th entry of our matrix.

Each diagonal: $((1,k),(2,k+1),(3,k+2),\dots)$ or
$((k,1),(k+1,2),(k+2,3),\dots)$ of a plane partition is an
(ordinary) partition ~-- a nonincreasing sequence of nonnegative
integers with only finitely many nonzero elements. A {\it strict}
partition is an ordinary partition with distinct positive elements.
The example given above has all diagonals strict and so it is a {\it
strict plane partition}~-- a plane partition whose diagonals are
strict partitions.

For a plane partition $\pi$ one defines the volume $|\pi|$ to
be the sum of all entries of the corresponding matrix, and $A(\pi)$
to be the number of ``white islands'' ~-- connected components of
white rhombi of the 3-dimensional representation of $\pi$.
\footnote{For the given example $|\pi|=35$ and $A(\pi)=7$.} Then
$2^{A(\pi)}$ is equal to the number of ways to color these islands
with two colors.

For a real number $q$, $0<q<1$, we define a probability
 measure $\mathfrak{M}_q$ on strict plane partitions as follows. If
$\pi$ is a strict plane partition we set $\text{Prob}(\pi)$ to be
proportional to $2^{A(\pi)}q^{|\pi|}$. The normalization constant is
the inverse of the partition function of these weights, which is
explicitly given by
\begin{propint} (Shifted MacMahon's formula)
$$
\sum_{\substack {\pi \text{ is a strict}\\ \text{plane partition}}
} 2^{A(\pi)}q^{|\pi|}=\prod_{n=1}^{\infty}\left(
\frac{1+q^n}{1-q^n}\right)^n.
$$
\end{propint}
This formula has appeared very recently in \cite{FW} and we were not
able to locate earlier references. This is a shifted version of the
famous MacMahon's formula:
$$
\sum_{\substack {\pi \text{ is a plane}\\ \text{partition}}
} q^{|\pi|}=\prod_{n=1}^{\infty}\left(
\frac{1}{1-q^n}\right)^n.
$$
A purely combinatorial proof of the shifted MacMahon's formula will appear in a subsequent publication.

The measure described above is a special case of the {\it shifted
Schur process}. This is a measure on sequences of strict (ordinary)
partitions. The idea came from an analogy with the {\it Schur
process} introduced in \cite{OR} that is a measure on sequences of
(ordinary) partitions. The Schur process is a generalization of the
{\it Schur measure} on partitions introduced earlier in \cite{O}.
The shifted Schur process we define is the generalization of the
{\it shifted Schur measure} that was introduced and studied in
\cite{TW} and \cite{Mat}.

The Schur process and the Schur measure have been extensively
studied in recent years and they have various applications, see e.g.
\cite{B}, \cite{BOk}, \cite{BOl}, \cite{BR},
\cite{IS}, \cite{J1}, \cite{J2}, \cite{OR2}, \cite{PS}.

In this paper we define the shifted Schur process and we derive the
formulas for its correlation functions in terms of Pfaffians of a
correlation kernel. We show that $\mathfrak{M}_q$ described above
can be seen as a special shifted Schur process. This allows us to
compute the correlation functions for this special case and further
to obtain their bulk scaling limit when $q \to 1$ (partitions become
large).

The shifted Schur process is defined using {\it skew Schur} $P$ and
$Q$ {\it functions}. These are symmetric functions that appear in
the theory of projective representations of the symmetric groups.

As all other symmetric functions, Schur $P_\lambda(x_1,x_2,\dots)$
function, where $\lambda$ is a strict partition,  is defined by a
sequence of polynomials $P_\lambda(x_1,x_2,\dots x_n)$, $n\in \bbN$,
with a property that $P_\lambda(x_1,x_2,\dots
x_n,0)=P_\lambda(x_1,x_2,\dots x_n)$.

For a partition $\lambda$ one defines the length $l(\lambda)$ to be
the number of nonzero elements. Let
$\lambda=(\lambda_1,\lambda_2,\dots)$ be a strict partition and
$n \geq l$, where $l$ is the length of $\lambda$. Then
$$
P_\lambda(x_1,\dots,x_n)=\sum_{w\in S_n/S_{n-l}}w\Big(x_1^{\lambda_1}\cdots x_l^{\lambda_{l}}\prod_{i=1}^{l}\prod_{j>i}\frac{x_i+x_j}{x_i-x_j}\Big),
$$
where $S_{n-l}$ acts on $x_{l+1},\dots,x_n$. Schur $Q_\lambda$ function is defined as $2^{l}P_\lambda$.

Both $\{P_\lambda:\lambda \;\text{strict}\}$ and
$\{Q_\lambda:\lambda\;\text{strict}\}$ are bases for
$\bbQ[p_1,p_3,p_5,\dots]$, where $p_r=\sum x_i^r$ is the $r$th power
sum. A scalar product in $\bbQ[p_1,p_3,p_5,\dots]$ is given with
$\langle P_\lambda,Q_\mu \rangle=\delta_{\lambda,\mu}$.

For strict partitions $\lambda=(\lambda_1,\lambda_2,\dots)$ and
$\mu=(\mu_1,\mu_2,\dots)$, skew Schur functions are defined by
$$
P_{\lambda / \mu }=
\begin{cases}
\sum_\nu\langle P_\lambda,Q_\lambda Q_\nu \rangle P_\nu & \lambda\supset \mu,\\
0&\text{otherwise},
\end{cases}
\;\;\;\;\;
\text{and}\;\;\;\;\;Q_{\lambda/\mu}=2^{l(\lambda)-l(\mu)}P_{\lambda/\mu},
$$
where $\lambda\supset\mu$ if $\lambda_i\geq\mu_i$ for every $i$.
Note that $P_{\lambda/\emptyset}=P_\lambda$ and
$Q_{\lambda/\emptyset}=Q_\lambda$.

This is just one of many ways to define skew Schur $P$ and $Q$
functions (see Chapter 3 of \cite{Mac}). We give another definition
in the paper that is more convenient for us.

We use $\Lambda$ to denote the algebra of symmetric functions. A
{\it specialization} of the algebra of symmetric functions is an
algebra homomorphism $\Lambda \to \bbC$. If $\rho$ is a
specialization and $f\in \Lambda$ then we use $f(\rho)$ to denote
the image of $f$ under $\rho$.

Let $\rho=(\rho_0^{+},\rho_1^{-},\rho_1^{+}, \ldots, \rho_T^{-})$ be
a finite sequence of specializations. For two sequences of strict
partitions $ \lambda=(\lambda^{1}, \lambda^{2},\ldots ,\lambda^{T})$
and $ \mu=(\mu^{1},\mu^{2}, \ldots ,\mu^{T-1})$ we define
$$W(\lambda, \mu)=Q_{\lambda^{1}} (\rho_0^{+})P_{\lambda^{1}/\mu^{1}}
({\rho_1^{-}}) Q_{\lambda^{2}/\mu^{1}} (\rho_1^{+}) \ldots
Q_{\lambda^{T}/\mu^{T-1}} ({\rho_{T-1}^{+}})P_{\lambda^{T}}
({\rho_T^{-}}).$$
Then $W(\lambda , \mu)=0$ unless
$$\emptyset \subset \lambda^{1} \supset \mu^{1} \subset \lambda^{2} \supset \mu^{2} \subset \ldots \supset \mu^{T-1} \subset
\lambda^{T} \supset \emptyset.$$

The shifted Schur process is a measure that to a sequence of strict
plane partitions $ \lambda=(\lambda^{1},\lambda^{2}, \ldots
,\lambda^{T})$ assigns
$${\rm Prob}(\lambda)=\frac{1}{Z(\rho)}\sum_{\mu}W(\lambda,\mu),$$
where $Z(\rho)$ is the partition function, and the sum goes over all
sequences of strict partitions $\mu=(\mu^{1},\mu^{2}, \ldots
,\mu^{T-1})$.


Let $X=\{(x_i,t_i):i=1,\dots,n \} \subset \bbN \times [1,2, \ldots
,T]$ and let $\lambda=(\lambda^{1}, \lambda^{2}, \dots, \lambda^T)$
be a sequence of strict partitions. We say that $X \subset \lambda$
if $x_i$ is a part (nonzero element) of the partition $\lambda^{t_i}$
for every $i=1,\dots,n$. Define the {\it correlation function} of
the shifted Schur process by
$$
\rho(X)={\text{Prob}(X\subset\lambda)}.
$$

The first main result of this paper is that the shifted Schur
process is a Pfaffian process, i.e. its correlation functions can be
expressed as Pfaffians of a certain kernel.

\begin{tA}
Let $X \subset \bbN \times [1,2, \ldots T]$ with $|X|=n$. The
correlation function has the form
$$\rho(X)=\Pf(M_X)$$
where $M_X$ is a skew-symmetric $2n \times 2n$ matrix
\begin{equation*}
M_X(i,j)=
\begin{cases} K_{x_i,x_j}(t_i,t_j) & \text{ $1 \leq i<j \leq n$,}\\
(-1)^{x_{j'}}K_{x_i,-x_j'}(t_i,t_{j'}) & \text{ $1 \leq i \leq n < j \leq 2n$,}\\
(-1)^{x_{i'}+x_{j'}}K_{-x_{i'},-x_{j'}}(t_{i'},t_{j'}) & \text{ $n <
i < j \leq 2n$,}
\end{cases}
\end{equation*}
where $i'=2n-i+1$ and $K_{x,y}(t_i,t_j)$ is the coefficient of
$z^xw^y$ in the formal power series expansion of
\begin{equation*}
\frac{z-w}{2(z+w)}J(z,t_i)J(w,t_j)
\end{equation*}
in the region $|z|>|w|$ if $t_i \geq t_j$ and $|z|<|w|$ if $t_i <
t_j$.

Here $J(z,t)$ is given with
\begin{equation*}
J(t,z)=\prod_{t \leq m }F(\rho_m^{-};z)\prod_{m \leq
t-1}F(\rho_m^{+};z^{-1}),
\end{equation*}
where $F(x;z)=\prod_i({1+x_iz})/({1-x_iz})$.
\end{tA}

Our approach is similar to that of \cite{OR}. It relies on two
tools. One is the Fock space associated to strict plane partitions
and the other one is a Wick type formula that yields a Pfaffian.

Theorem A can be used to obtain the correlation functions for
$\mathfrak{M}_q$ that we introduced above. In that case
\begin{equation*}
J_q(t,z)=
\begin{cases}
\displaystyle\frac{(q^{1/2} z^{-1};q)_\infty(-q^{t+1/2}z;q)_\infty}{(-q^{1/2} z^{-1};q)_\infty(q^{t+1/2}z;q)_\infty}&t\geq0,\\
\\
 \displaystyle\frac{(-q^{1/2}z;q)_\infty(q^{-t+1/2} z^{-1};q)_\infty}{(q^{1/2}z;q)_\infty(-q^{-t+1/2} z^{-1};q)_\infty}&t<0,
\end{cases}
\end{equation*}
where
$$
(z;q)_\infty=\prod_{n=0}^{\infty}(1-q^nz)
$$
is the quantum dilogarithm function. We use the Pfaffian formula to
study the bulk scaling limit of the correlation functions when $q
\to 1$. We scale the coordinates of strict plane partitions by
$r=\log q$. This scaling assures that the scaled volume of strict
plane partitions tends to a constant.\footnote{The constant is equal
to $7\zeta(3)/4$.}

Before giving the statement, we need to say that a strict plane
partition is uniquely represented by a point configuration (subset) in
$$
\frak{X}=\left\{(t,x)\in \bbZ \times \bbZ
\left|\right.x>0\right\}
$$
as follows: $(i,j,\pi(i,j)) \to (j-i,\pi(i,j))$, where
$\pi(i,j)$ is the $(i,j)$th entry of $\pi$.

Let
$\gamma^+_{R,\theta}$, respectively $\gamma^-_{R,\theta}$, be the
counterclockwise, respectively clockwise, arc of $|z|=R$ from
$Re^{-i\theta}$ to $Re^{i\theta}$.
\begin{tB}
Let $X=\{(t_i,x_i):i=1, \dots, n\}\subset \frak{X}$ be such that
$$
rt_i \to \tau, \;\;\; rx_i \to \chi \;\;\; \text{as } r \to +0,
$$
$$ t_i-t_j= \Delta t_{ij}=\text{const}, \;\;\; x_i-x_j= \Delta x_{ij}=
\text{const}.
$$
a) If $\chi > 0$ then
$$
\lim_{r \to +0}\rho(X) = \det[K(i,j)]_{i,j=1}^n,
$$
where
$$
K(i,j)=\frac{1}{2\pi i} \int_{\gamma_{R,\theta} ^\pm} \left(
\frac{1-z}{1+z} \right)^{\Delta t_{ij}} \frac{1}{z^{\Delta
x_{ij}+1}}dz,
$$
where we choose $\gamma_{R,\theta} ^+$ if $\Delta t_{ij} \geq 0$ and
$\gamma_{R,\theta} ^-$ otherwise, where $R=e^{-|\tau|/2}$ and
\begin{equation*}
\theta=\begin{cases}\arccos\displaystyle\frac{(e^{|\tau|}+1)(e^{\chi}-1)}{2e^{|\tau|/2}{(e^{\chi}+1)}},
&\displaystyle\frac{(e^{|\tau|}+1)(e^{\chi}-1)}{2e^{|\tau|/2}{(e^{\chi}+1)}}\leq 1,\\
0,&\text{otherwise,}
\end{cases}
\end{equation*}

b) If $\chi=0$ and in addition to the above conditions we
assume
$$
x_i=\text{const}
$$
then
$$
\lim_{r \to +0}\rho(X) = \Pf [M(i,j)]_{i,j=1}^{2n},
$$
where $M$ is a skew symmetric matrix given by
$$
M(i,j)=\begin{cases} \displaystyle \frac{(-1)^{x_j}}{2\pi i}
\int_{\gamma_{R,\theta} ^\pm}
\left(\frac{1-z}{1+z} \right)^{\Delta t_{ij}}\frac{dz}{z^{x_i+x_j+1}}& \text{ $1 \leq i<j \leq n$,}\\
 \displaystyle \frac{1}{2\pi i}
\int_{\gamma_{R,\theta} ^\pm}
\left(\frac{1-z}{1+z} \right)^{\Delta t_{ij'}}\frac{dz}{z^{x_i-x_{j'}+1}}&\text{ $1 \leq i \leq n < j \leq 2n$,}\\
\displaystyle \frac{(-1)^{x_{i'}}}{2\pi i} \int_{\gamma_{R,\theta}
^\pm} \left(\frac{1-z}{1+z} \right)^{\Delta
t_{i'j'}}\frac{dz}{z^{-(x_{i'}+x_{j'})+1}}\ & \text{ $n < i < j \leq
2n$,}
\end{cases}
$$
where $i'=2n-i+1$ and we choose $\gamma _{R,\theta} ^+$ if $\Delta
t_{ij} \geq 0$ and $\gamma _{R,\theta} ^-$ otherwise, where
$R=e^{-|\tau|/2}$ and $\theta=\pi/2$.
\end{tB}


For the equal time configuration (points on the same vertical line)
we get the discrete sine kernel and thus, the kernel of the theorem
is an  extension of the sine kernel. This extension has appeared  in
\cite{B}, but there it did not come from a ``physical'' problem.

The theorem above allows us to obtain the limit shape of large
strict plane partitions distributed according to $\mathfrak{M}_q$,
but we do not prove its existence. The limit shape is parameterized
on the domain representing a half of the amoeba\footnote{The amoeba
of a polynomial $P(z,w)$
 is $$ \{(\xi,\omega)=(\log|z|,\log|w|)\in
\bbR^2\;|\;(z,w)\in(\bbC\backslash \{0\})^2, \; P(z,w)=0\}. $$} of
the polynomial $P(z,w)=-1+z+w+zw$ (see Section 4 for details).

In the proof of Theorem B we use the saddle point analysis. We
deform contours of integral that defines elements of the correlation
kernel in a such a way that it splits into an integral that vanishes
when $q\to 1$ and another nonvanishing integral that comes as a
residue.

The paper is organized as follows. In Section 2 we introduce the
shifted Schur process and compute its correlation function (Theorem
A). In Section 3 we introduce the strict plane partitions and the
measure $\mathfrak{M}_q$. We prove the shifted MacMahon's formula
and use Theorem A to obtain the correlation functions for this
measure. In Section 4 we compute the scaling limit (Theorem B).
Section 5 is an appendix  where we give a summary of
definitions we use and explain the Fock space formalism associated
to strict plane partitions.

\begin{acknowledgement}
This work is a part of my doctoral dissertation at California
Institute of Technology and I thank my advisor Alexei Borodin for
all his help and guidance. Also, I thank Percy Deift for helpful discussions.
\end{acknowledgement}
\section{The shifted Schur process} \lb{s2}

\subsection{The measure} \lb{s2.1}

Recall that  a nonincreasing sequence $\lambda=(\lambda_1,
\lambda_2, \dots)$ of nonnegative integers with a finite number of
parts (nonzero elements) is called a partition. A partition is called strict
if all parts are distinct. More information on strict
partitions can be found in \cite{Mac}, \cite{Mat}. Some
results from these references that we are going to use are
summarized in Appendix to this paper.

The shifted Schur process is a measure on a space of sequences of
strict partitions. This measure depends on a finite sequence
$\rho=(\rho_0^{+},\rho_1^{-},\rho_1^{+}, \ldots, \rho_T^{-})$ of
specializations of the algebra $\Lambda$ of symmetric functions
\footnote{A specialization of $\Lambda$ is an algebra homomorphism
$\Lambda \to \bbC$.}.


Let $ \lambda=(\lambda^{1},\lambda^{2}, \ldots ,\lambda^{T})$ and $
\mu=(\mu^{1},\mu^{2}, \ldots ,\mu^{T-1})$ be two sequences of strict
partitions. Set
$$W(\lambda, \mu)=Q_{\lambda^{1}} (\rho_0^{+})P_{\lambda^{1}/\mu^{1}}
({\rho_1^{-}}) Q_{\lambda^{2}/\mu^{1}} (\rho_1^{+}) \ldots
Q_{\lambda^{T}/\mu^{T-1}} ({\rho_{T-1}^{+}})P_{\lambda^{T}}
({\rho_T^{-}}).$$

Here $P_{\lambda / \mu} (\rho)$'s and $Q_{\lambda / \mu} (\rho)$'s
denote the skew Schur $P$ and $Q$-functions, see Appendix.

Note that $W(\lambda , \mu)=0$ unless
$$\emptyset \subset \lambda^{1} \supset \mu^{1} \subset \lambda^{2} \supset \mu^{2} \subset \ldots \supset \mu^{T-1} \subset
\lambda^{T} \supset \emptyset.$$

\begin{proposition}\lb{Z}
The sum of the weights $W(\lambda, \mu)$ over all sequences of
strict partitions $ \lambda=(\lambda^{1},\lambda^{2}, \ldots
,\lambda^{T})$ and $ \mu=(\mu^{1},\mu^{2}, \ldots ,\mu^{T-1})$ is
equal to
\begin{equation} \lb{FormulaForZ}
Z(\rho)=\prod_{0 \leq i < j \leq T}H(\rho_i^{+},\rho_j^{-}),
\end{equation}
where
$$H(\rho_i^{+},\rho_j^{-})=\sum_{\lambda \; {\rm strict}}Q_{\lambda}(\rho_i^{+})P_{\lambda}(\rho_j^{-}).$$
\end{proposition}
We give two proofs of this statement.
\begin{proof} {\bf 1.}
From (\ref{G-}) and (\ref{G+}) it follows that
\begin{equation*}
Z(\rho)=\sum_{\lambda,\mu}W(\lambda,\mu)= \langle
\Gamma_{+}(\rho_T^{-})\Gamma_{-}(\rho_{T-1}^{+}) \cdots
\Gamma_{-}(\rho_1^{+})\Gamma_{+}(\rho_1^{-})\Gamma_{-}(\rho_0^{+})
v_\emptyset, v_\emptyset
 \rangle.
\end{equation*}
We can move all $\Gamma_{+}$'s to the right and all $\Gamma_{-}$'s
to the left using (\ref{G+G-}). We obtain
\begin{equation*}
Z(\rho)= \prod_{0 \leq i < j \leq T}H(\rho_i^{+},\rho_j^{-})\langle
\Gamma_{-}(\rho_{T-1}^{+}) \cdots \Gamma_{-}(\rho_{0}^{+})
\Gamma_{+}(\rho_T^{-}) \cdots \Gamma_{+}(\rho_1^{-}) v_\emptyset,
v_\emptyset
 \rangle.
\end{equation*}
Then (\ref{G+v0}) implies (\ref{FormulaForZ}).
\end{proof}

\begin{proof} {\bf 2.}
We use Proposition \ref{QPQP}. The idea is the same as in
Proposition 2.1 of \cite{BR}. The proof goes by induction on $T$.

Using the formula from Proposition \ref{QPQP} we substitute sums
over $\lambda^{i}$'s with sums over $\tau^{i-1}$'s. This gives
$$\prod_{i=0}^{T-1}H({\rho_i^{+}},{\rho_{i+1}^{-}}) \sum_{\mu, \tau}Q_{\mu^{1}} (\rho_0^{+})P_{\mu^{1}/\tau^{1}}
({\rho_2^{-}}) Q_{\mu^{2}/\tau^{1}} (\rho_1^{+}) \ldots
P_{\mu^{T-1}} ({\rho_{T}^{-}}).$$ This is the sum of $W(\mu,\tau)$
with  $ \mu=(\mu^{1}, \ldots ,\mu^{T-1})$ and $ \tau=(\tau^{1},
\ldots ,\tau^{T-2})$. Inductively, we obtain (\ref{FormulaForZ}).
\end{proof}

\begin{definition}
The shifted Schur process is a measure on the space of finite sequences
that to $\lambda=(\lambda^{1},\lambda^{2}, \ldots ,\lambda^{T})$ assigns
$${\rm Prob}(\lambda)=\frac{1}{Z(\rho)}\sum_{\mu}W(\lambda,\mu),$$
where the sum goes over all $
\mu=(\mu^{1},\mu^{2}, \ldots ,\mu^{T-1})$.
\end{definition}

\subsection{Correlation functions} \lb{s2.2}

Our aim is to compute the correlation functions for the shifted
Schur process.

Let $X=\{(x_i,t_i):i=1,\dots,n \} \subset \bbN \times [1,2, \ldots
,T]$ and let $\lambda=(\lambda^{1}, \lambda^{2}, \dots, \lambda^T)$
be a sequence of strict partitions. We will say that $X \subset
\lambda$ if $x_i$ is a part of the partition $\lambda^{t_i}$ for every
$i=1,\dots,n$.

\begin{definition}Let $X \subset \bbN \times [1,2, \ldots T]$. Then the correlation function of the shifted Schur process
corresponding to $X$ is
$$\rho(X)={\rm Prob}(X \subset \lambda).$$
\end{definition}

We are going to show that the shifted Schur process is a Pfaffian
process in the sense of \cite{BR}, that is, its correlation
functions are Pfaffians of submatrices of a fixed matrix called the
correlation kernel. This is stated in the following theorem.
\begin{theorem} \lb{CorFun}
Let $X \subset \bbN \times [1,2, \ldots T]$ with $|X|=n$. The
correlation function is given with
$$\rho(X)=\Pf(M_X)$$
where $M_X$ is a skew-symmetric $2n \times 2n$ matrix
\begin{equation*}
M_X(i,j)=
\begin{cases} K_{x_i,x_j}(t_i,t_j) & \text{ $1 \leq i<j \leq n$,}\\
(-1)^{x_{j'}}K_{x_i,-x_j'}(t_i,t_{j'}) & \text{ $1 \leq i \leq n < j \leq 2n$,}\\
(-1)^{x_{i'}+x_{j'}}K_{-x_{i'},-x_{j'}}(t_{i'},t_{j'}) & \text{ $n <
i < j \leq 2n$,}
\end{cases}
\end{equation*}
where $i'=2n-i+1,\,j'=2n-j+1$ and $K_{x,y}(t_i,t_j)$ is the
coefficient of $z^xw^y$ in the formal power series expansion of
\begin{equation*}
K((z,t_i),(w,t_j)):=\sum_{x,y \in
\bbZ}K_{x,y}(t_i,t_j)z^xw^y=\frac{z-w}{2(z+w)}J(z,t_i)J(w,t_j)
\end{equation*}
in the region $|z|>|w|$ if $t_i \geq t_j$ and $|z|<|w|$ if $t_i <
t_j$.

Here $J(z,t)$ is given with
\begin{equation} \lb{J}
J(t,z)=\prod_{t \leq m }F(\rho_m^{-};z)\prod_{m \leq
t-1}F(\rho_m^{+};z^{-1}),
\end{equation}
where $F$ is defined with (\ref{Q}).
\end{theorem}

\begin{proof}
The proof consists of two parts. In the first part we express the
correlation function via the operators $\Gamma_{+}$ and $\Gamma_{-}$
(see Appendix). In the second part we use a Wick type formula to
obtain the Pfaffian.

Let $\rho_T^{+}=\rho_0^{-}=0$ and let $t_0=0$.

First, we assume $1 \leq t_1 \leq t_2 \leq \cdots \leq t_n \leq T$. Using formulas (\ref{G-}), (\ref{G+}) and (\ref{PsiPsi*}) we get that the
correlation function is
\begin{equation*}
\frac{1}{Z(\rho)} \biggr< \prod_{m=t_n}^{T}
\Gamma_{-}(\rho_m^{+})\Gamma_{+}(\rho_m^{-}) \prod_{i=1}^n
\Big(2\psi_{x_i}\psi_{x_i}^{*} \prod_{m={t_{i-1}}}^{t_i-1}
\Gamma_{-}(\rho_{m}^{+})\Gamma_{+}(\rho_m^{-})\Big)
v_{\emptyset},v_{\emptyset}  \biggl>,
\end{equation*}
where the products of the operators should be read from right to
left in the increasing time order.

Thus, the correlation function is equal to
$\prod_{i=1}^{n}(-1)^{x_{i}}$ times the  coefficient of
$\prod_{i=1}^{n} u_i^{x_i}{v_i}^{-x_i}$ in the formal power series
\begin{equation*}
\frac{1}{Z(\rho)}\biggr< \prod_{m=t_n}^{T}
\Gamma_{-}(\rho_m^{+})\Gamma_{+}(\rho_m^{-}) \prod_{i=1}^n
\Big(2\psi(u_i)\psi(v_i) \prod_{m={t_{i-1}}}^{t_i-1}
\Gamma_{-}(\rho_{m}^{+})\Gamma_{+}(\rho_m^{-})
\Big)v_{\emptyset},v_{\emptyset} \biggl>,
\end{equation*}
where $\psi$ is given by (\ref{defpsi}).

We use formulas (\ref{G+G-}) and (\ref{GFunct}) to put all
$\Gamma_{-}$'s on the left and $\Gamma_{+}$'s on the right side of
$\prod_{i=1}^n 2\psi(u_i)\psi(v_i)$. Since
$\Gamma_{+}v_{\emptyset}=v_{\emptyset}$, see (\ref{G+v0}), we obtain
\begin{equation} \lb{JPsi} \prod_{i=1}^n
(J(t_i,u_i)J(t_i,v_i)) \biggr< \prod_{i=1}^n
2\psi(u_i)\psi(v_i)v_{\emptyset},v_{\emptyset} \biggl>.
\end{equation}
Thus, if $1 \leq t_1 \leq t_2 \leq \cdots \leq t_n \leq T$ then the
correlation function $\rho(X)$ is equal to
$\prod_{i=1}^{n}(-1)^{x_{i}}$ times the  coefficient of
$\prod_{i=1}^{n} u_i^{x_i}{v_i}^{-x_i}$ in the formal power series
(\ref{JPsi}).

Now, more generally, let $\pi \in S_n$ such that $1 \leq t_{\pi(1)}
\leq t_{\pi(2)} \leq \cdots \leq t_{\pi(n)} \leq T$, then the
correlation function $\rho(X)$ is equal to
$\prod_{i=1}^{n}(-1)^{x_{i}}$ times the  coefficient of
$\prod_{i=1}^{n} u_i^{x_i}{v_i}^{-x_i}$ in the formal power series
\begin{equation} \lb{JPsiGen}
\prod_{i=1}^n (J(t_i,u_i)J(t_i,v_i)) \biggr< \prod_{i=1}^n
2\psi(u_{\pi(i)})\psi(v_{\pi(i)})v_{\emptyset},v_{\emptyset} \biggl>.
\end{equation}
The above inner product is computed in Lemma \ref{PfaffianPsi1} and
Lemma \ref{PfaffianPsi2}, that will be proved below. Then by Lemma \ref{PfaffianPsi2}  we
obtain that in the region $|u_{\pi(n)}|>|v_{\pi(n)}|> \cdots
>|u_{\pi(1)}|>|v_{\pi(1)}|$ expression (\ref{JPsiGen}) is equal to
$\Pf(A)$ with
\begin{equation*}
A(i,j)=
\begin{cases} K((u_i,t_i),(u_j,t_j)) & \text{ $1 \leq i<j \leq n$,}\\
K((u_i,t_i),(v_j',t_j')) & \text{ $1 \leq i \leq n < j \leq 2n$,}\\
K((v_i',t_i'),(v_j',t_j')) & \text{ $n< i < j \leq 2n$,}
\end{cases}
\end{equation*}
where $i'=2n-i+1,\,j'=2n-j+1$ and $K$ is as above.

Let $(y_1,\dots,y_{2n})=(x_1,\dots,x_n,-x_n,\dots,-x_1)$ and
$(z_1,\dots,z_{2n})=(u_1,\dots,u_n,v_n,\dots,v_1)$. By the
definition of the Pfaffian,
\begin{eqnarray*}
\Pf(M_X)=\prod_{i=1}^{n}(-1)^{x_i}\sum_\alpha
{\text{sgn}(\alpha)}\prod_sK_{y_{i_s},y_{j_s}},
\end{eqnarray*}
where the sum is taken over all permutations
$\alpha=\left(\begin{array}{ccccc}1&2&\cdots&2n-1&2n\\i_1&j_1&\cdots
&i_n&j_n\end{array}\right)$, such that $i_1<\cdots<i_n$ and
$i_s<j_s$ for every $s$.

Also,
\begin{eqnarray*}
\Pf(A)&=&\sum_\alpha \text{sgn}(\alpha)\prod_s
K({z_{i_s},z_{j_s}})\\
&=&\sum_{y_1,\dots,y_{2n}\in \bbZ}\sum_\alpha
\text{sgn}(\alpha)\prod_s
K_{y_{i_s},y_{j_s}}\prod_{i=1}^{2n}z_i^{y_i}.
\end{eqnarray*}
Finally, since
\begin{equation*}
\rho(X)=\prod_{i=1}^{n}(-1)^{x_i}[\prod_{i=1}^{n}
u_i^{x_i}{v_i}^{-x_i}:\Pf(A)]\\
\end{equation*}
we get
\begin{equation*}
\rho(X)=\Pf(M_X).
\end{equation*}
\end{proof}

\begin{lemma}\lb{PfaffianCoefficients}
\begin{equation}\lb{lrlemma}
\langle \psi_{k_{2n}}\cdots
\psi_{k_1}v_\emptyset,v_\emptyset\rangle=
\sum_{i=2}^{2n}(-1)^i\langle\psi_{k_{2n}}\cdots \hat
\psi_{k_i}\cdots\psi_{k_2}v_\emptyset,v_\emptyset
\rangle\langle\psi_{k_i}\psi_{k_1}v_\emptyset,v_\emptyset\rangle
\end{equation}
Equivalently,
$$
\langle \psi_{k_{2n}}\cdots
\psi_{k_1}v_\emptyset,v_\emptyset\rangle=\Pf[A_{ij}]_{i,j=1}^{2n},
$$
where $A$ is a skew symmetric matrix with
$A_{ij}=\langle\psi_{k_j}\psi_{k_i}v_\emptyset,v_\emptyset\rangle$ for $i<j$.
\end{lemma}
\begin{proof}
Throughout the proof we use $\langle\cdots
v_\emptyset,v_\emptyset\rangle=\langle\cdots \rangle$ to shorten the
notation.

We can simplify the proof if we use the fact that the operators
$\psi_k$ and $\psi_k^*$ add, respectively remove $e_k$, but the
proof we are going to give will apply to a more general case, namely,
we consider any $ \psi_k$'s that satisfy
\begin{equation} \lb{vacuum}
\psi_k v_{\emptyset}=0,\;\;\;k<0,
\end{equation}
\begin{equation}\lb{psinula}
\psi_0v_\emptyset=a_0v_{\emptyset},
\end{equation}
\begin{equation}\lb{psiadjoint}
\psi_k^{*}=b_k\psi_{-k},
\end{equation}
\begin{equation}\lb{commute}
\psi_k\psi_l+\psi_l\psi_k=c_k\delta_{k,-l},
\end{equation}
for some constants $a_0,b_k$ and $c_k$. These properties are
satisfied for $\psi_k$ we are using (see Appendix).

The properties (\ref{vacuum})-(\ref{commute}) immediately imply that
\begin{equation}\lb{psikpsik}
\psi_k\psi_k=0,\;\;\;k\neq0
\end{equation}
and
\begin{equation} \lb{psikpsil}
\langle\psi_k\psi_l\rangle=0,\;\;\;k\neq-l
\end{equation}
because for $k\neq-l$
$$
\langle\psi_k\psi_l\rangle=
\begin{cases}
\langle\psi_k\psi_lv_\emptyset,v_\emptyset\rangle&l<0,\\
-b_l\langle\psi_kv_\emptyset,\psi_{-l}v_\emptyset\rangle&l>0,\\
a_0\langle\psi_kv_\emptyset,v_\emptyset\rangle&l=0.\\
\end{cases}
$$

Observe that if $k_1,\dots,k_{2n-1}$ are all different from $0$ then
\begin{equation}\lb{nepar}
\langle \psi_{k_{2n-1}}\cdots \psi_{k_1}\rangle=0.
\end{equation}
This is true because there is $k>0$ such that $(k_1,\dots,k_{2n-1})$
has odd number of elements whose absolute value is $k$. Let those be
$k_{j_1},\dots ,k_{j_{2s-1}}$. If there exist $k_{j_i}$ and
$k_{j_{i+1}}$ both equal either to $k$ or $-k$ then
$\psi_{k_{2n-1}}\cdots \psi_{k_1}=0$ by (\ref{commute}) and
(\ref{psikpsik}). Otherwise, $(k_{j_1},\dots ,k_{j_{2s-1}})$ is
either $(k,-k,\dots ,-k,k)$ or $(-k,k,\dots ,k,-k)$. In the first
case we move $\psi_k$ to the left and use (\ref{vacuum}) and
(\ref{psiadjoint}) and in the second we move $\psi_{-k}$ to the
right and use (\ref{vacuum}).

Also, observe that if $(k_n,\dots,k_1)$ contains $t$ zeros and if
those appear on places $j_t,\dots,j_1$ then
\begin{equation}
\langle\psi_{k_n}\cdots\psi_{k_1}\rangle=a_0^t\langle\psi_{k_n}\cdots\hat\psi_{k_{j_t}}\cdots\hat\psi_{k_{j_1}}\cdots\psi_{k_1}\rangle(-1)^{\sum_{m=1}^t
j_m-m}.
\end{equation}
This follows from (\ref{psinula}) and (\ref{commute}) by moving
$\psi_{k_{j_m}}$ to the $m$th place from the right.

Now, we proceed to the proof of the lemma. We use induction on $n$.
We need to show that $L(k_{2n},\dots,k_1)=R(k_{2n},\dots,k_1)$ where
$L$ and $R$ are the left, respectively right hand side of (\ref{lrlemma}). Obviously it is true for
$n=1$. We show it is true for $n$.

If $k_1<0$ then $L=R=0$ by (\ref{vacuum}).

If $k_1=0$ then let $t$ be the number of zeros in $k_{2n},\dots,k_1$
and let those appear on places $j_t,\dots,j_1=1.$ Then
$$
L=a_0^t\langle\psi_{k_{2n}}\cdots\hat\psi_{k_{j_t}}\cdots\hat\psi_{k_{j_1}}\cdots\psi_{k_1}\rangle\prod_{m=1}^t(-1)^{
j_m-m},
$$
while
\begin{eqnarray*}
R&=&\sum_{i=2}^t(-1)^{j_i}\langle\psi_{k_{2n}}\cdots\hat\psi_{k_{j_i}}\cdots\psi_{k_2}\rangle
\langle\psi_{k_{j_i}}\psi_{k_1}\rangle\\
&=&\sum_{i=2}^t(-1)^{j_i}a_0^t\langle\psi_{k_{2n}}\cdots\hat\psi_{k_{j_t}}\cdots\hat\psi_{k_{j_1}}\cdots\psi_{k_1}\rangle
\prod_{\substack{ m=2 \\m \neq i}}^t(-1)^{j_m-m}\\
&=&a_0^t\langle\psi_{k_{2n}}\cdots\hat\psi_{k_{j_t}}\cdots\hat\psi_{k_{j_1}}\cdots\psi_{k_1}\rangle\prod_{m=1}^t(-1)^{
j_m-m}\sum_{i=2}^t(-1)^i.
\end{eqnarray*}
If $t$ is even then $L=R$. If $t$ is odd then $2n-t$ is odd then
$\langle\psi_{k_{2n}}\cdots\hat\psi_{k_{j_t}}\cdots\hat\psi_{k_{j_1}}\cdots\psi_{k_1}\rangle=0$
by (\ref{nepar}) and thus $L=R=0$.

Finally, we can assume $k_1 > 0$. If $k_i\neq-k_1$ for every $i \in
\{2,\dots,n\}$ then
$$
L(k_{2n},\dots,k_1)=R(k_{2n},\dots,k_1)=0.
$$
This is true because $R=0$ by (\ref{psikpsil}) and $L=0$ since it is
possible to move $\psi_{k_1}$ to the left and then use
(\ref{vacuum}) and (\ref{psiadjoint}).

So, we need to show that $L=R$ if $k_1 > 0$  and there exists $i \in
\{2,\dots,n\}$ such that $k_i=-k_1$.

First, we assume $k_2=-k_1$. Then
$$
L=\langle\psi_{k_{2n}}\cdots\psi_{k_3}\rangle
\langle\psi_{k_2}\psi_{k_1}\rangle,
$$
because
$\psi_{-k}\psi_{k}v_\emptyset=\langle\psi_{-k}\psi_{k}\rangle
v_{\emptyset}$. On the other hand
$$
R=\langle\psi_{k_{2n}}\cdots\psi_{k_3}\rangle
\langle\psi_{k_2}\psi_{k_1}\rangle,
$$
because for every $i>2$ we have that $\langle\psi_{k_{2n}}\cdots
\hat \psi_{k_i}\cdots\psi_{k_2} \rangle=0$ by (\ref{vacuum}). Thus,
$L=R.$

To conclude the proof we show that
\begin{equation}\lb{implication}
\begin{array}{ccc}
L(k_{2n},\dots,k_s,k_{s-1},\dots,
k_1)&=&R(k_{2n},\dots,k_s,k_{s-1},\dots, k_1)\\
&\Downarrow&\\
L(k_{2n},\dots,k_{s-1},k_{s},\dots,
k_1)&=&R(k_{2n},\dots,k_{s-1},k_{s},\dots, k_1).\\
\end{array}
\end{equation}
Then we use this together with the fact that there is $i$ such that
$k_i=-k_1$ and that $L=R$ if $i=2$ to prove the induction step.
\begin{eqnarray*}
&&L(k_{2n},\dots,k_{s-1},k_{s},\dots,
k_1)=\langle\psi_{k_{2n}}\cdots\psi_{k_{s-1}}\psi_{k_{s}}\cdots\psi_{k_1}\rangle=\\
&&=-\langle\psi_{k_{2n}}\cdots\psi_{k_{s}}\psi_{k_{s-1}}\cdots\psi_{k_1}\rangle+c_{k_s}\delta_{k_s,-k_{s-1}}\langle\psi_{k_{2n}}\cdots\hat\psi_{k_{s}}\hat\psi_{k_{s-1}}\cdots\psi_{k_1}\rangle\\
&&=-L(k_{2n},\dots,k_s,k_{s-1},\dots,
k_1)+c_{k_s}\delta_{k_s,-k_{s-1}}\langle\psi_{k_{2n}}\cdots\hat\psi_{k_{s}}\hat\psi_{k_{s-1}}\cdots\psi_{k_1}\rangle
\end{eqnarray*}
\begin{eqnarray*}
&&R(k_{2n},\dots,k_{s-1},k_{s},\dots, k_1)=\\
&=&\sum_{\substack{i=2\\i \neq
s,s-1}}^{2n}(-1)^i\left[-\langle\psi_{k_{2n}}... \hat
\psi_{k_i}...\psi_{k_1}\rangle+c_{k_s}\delta_{k_s,-k_{s-1}}\langle\psi_{k_{2n}}...
\hat \psi_{k_i}...\hat\psi_{k_s}\hat\psi_{k_{s-1}}...\psi_{k_1}
\rangle\right]\langle\psi_{k_i}\psi_{k_1}\rangle\\
&&+(-1)^{s-1}\langle\psi_{k_{2n}}... \hat
\psi_{k_s}...\psi_{k_1}\rangle\langle\psi_{k_{s}}\psi_{k_1}\rangle+(-1)^s\langle\psi_{k_{2n}}...
\hat
\psi_{k_{s-1}}...\psi_{k_1}\rangle\langle\psi_{k_{s-1}}\psi_{k_1}\rangle\\
&=&-R(k_{2n},\dots,k_s,k_{s-1},\dots,
k_1)\\
&&+c_{k_s}\delta_{k_s,-k_{s-1}}\sum_{\substack{i=2\\i \neq
s,s-1}}^{2n}(-1)^i\langle\psi_{k_{2n}}... \hat
\psi_{k_i}...\hat\psi_{k_s}\hat\psi_{k_{s-1}}...\psi_{k_1}
\rangle\langle\psi_{k_i}\psi_{k_1}\rangle.
\end{eqnarray*}
Since, by the inductive hypothesis
$$
\langle\psi_{k_{2n}}...\hat\psi_{k_{s}}\hat\psi_{k_{s-1}}...\psi_{k_1}\rangle=
\sum_{\substack{i=2\\i \neq
s,s-1}}^{2n}(-1)^i\langle\psi_{k_{2n}}... \hat
\psi_{k_i}...\hat\psi_{k_s}\hat\psi_{k_{s-1}}...\psi_{k_1}
\rangle\langle\psi_{k_i}\psi_{k_1}\rangle
$$
we conclude that (\ref{implication}) holds.

\end{proof}

\begin{lemma} \lb{PfaffianPsi1}
Let $\pi \in S_n$.
\begin{equation*}
\biggl< \prod_{i=1}^n
\psi(u_{\pi(i)})\psi(v_{\pi(i)})v_{\emptyset},v_{\emptyset}
\biggr>=\Pf(\Psi_\pi),
\end{equation*}
where $\Psi_\pi$ is a skew-symmetric $2n$-matrix given with
\begin{equation*}
\Psi_\pi(i,j)=
\begin{cases} \psi_\pi((u_i,i),(u_j,j)) & \text{ $1 \leq i<j \leq n$,}\\
\psi_\pi((u_i,i),(v_j',j')) & \text{ $1 \leq i \leq n < j \leq 2n$,}\\
\psi_\pi((v_i',i'),(v_j',j')) & \text{ $n< i< j \leq 2n$,}
\end{cases}
\end{equation*}
where $i'=2n-i+1,\,j'=2n-j+1$ and
\begin{equation*}
\psi_\pi((z,i),(w,j))=
\begin{cases}
\langle \psi(z)\psi(w)v_{\emptyset},v_{\emptyset} \rangle & \text{ $\pi^{-1}(i) \geq \pi^{-1}(j)$,}\\
-\langle \psi(w)\psi(z)v_{\emptyset},v_{\emptyset} \rangle & \text{ $\pi^{-1}(i)<\pi^{-1}(j)$}.
\end{cases}
\end{equation*}
\end{lemma}

\begin{proof}
First we show that the statement is true for $\pi=id$. From Lemma
\ref{PfaffianCoefficients} we have
\begin{eqnarray*}
&&\langle \psi(u_{n})\psi(v_{n}) \cdots \psi(u_{1})\psi(v_{1})
v_{\emptyset},v_{\emptyset} \rangle=\langle \psi_{k_{2n}}\cdots
\psi_{k_1}v_\emptyset,v_\emptyset\rangle \prod_{i=1}^{n} u_iv_i\\
&&= \sum_{i=2}^{2n}(-1)^i\langle\psi_{k_{2n}}\cdots \hat
\psi_{k_i}\cdots\psi_{k_1}v_\emptyset,v_\emptyset
\rangle\langle\psi_{k_i}\psi_{k_1}v_\emptyset,v_\emptyset\rangle\prod_{i=1}^{n} u_iv_i\\
&&=\sum_{i=1}^{n} \langle \psi(u_{n})\psi(v_{n}) \cdots
\widehat{\psi(u_i)} \cdots
\psi(u_1)v_{\emptyset},v_{\emptyset}\rangle \langle
\psi(u_i)\psi(v_1)
v_{\emptyset},v_{\emptyset} \rangle\\
&&-\sum_{i=2}^{n} \langle \psi(u_{n})\psi(v_{n}) \cdots
\widehat{\psi(v_i)} \cdots
\psi(u_1)v_{\emptyset},v_{\emptyset}\rangle \langle
\psi(v_i)\psi(v_1) v_{\emptyset},v_{\emptyset} \rangle.
\end{eqnarray*}
Then by the expansion formula for the Pfaffian we get that
$$
\langle \psi(u_{n})\psi(v_{n}) \cdots \psi(u_{1})\psi(v_{1})
v_{\emptyset},v_{\emptyset} \rangle=\Pf(A(u_n,v_n,\dots,u_1,v_1)),
$$
where
$A(u_n,v_n, \ldots,u_1,v_1)$ is a skew symmetric $2n \times 2n$
matrix
\begin{equation*}
\left[
\begin{array}{ccccc}
0  & a(u_{n},v_{n})  & \dots & a(u_{n},u_{1})  & a(u_{n},v_{1})\\
-a(u_{n},v_{n})  & 0 & \dots & a(v_{n},u_{1})  & a(v_{n},v_{1})\\
\vdots&&\ \ddots && \dots \\
-a(u_{n},u_{1})  & -a(v_{n},u_{1}) & \dots & 0  & a(u_{1},v_{1})\\
 -a(u_{n},v_{1}) & -a(v_{n},v_{1})  & \dots & -a(u_{1},v_{1})  & 0\\
\end{array} \right],
\end{equation*}
with
\begin{equation*}
a(z,w)=\langle \psi(z)\psi(w) v_\emptyset, v_\emptyset \rangle.
\end{equation*}
Thus, the columns and rows of $A$ appear ``in order''
$u_n,v_n,\dots,u_1,v_1$. Rearrange these columns and columns ``in
order'' $u_1,u_2,\dots,u_n,v_n,\dots,v_2,v_1$. Let $B$ be the new
matrix. Since the number of switches we have to make to do this
rearrangement is equal to $n(n-1)$ and that is even, we have that
$\Pf(A)=\Pf(B)$ and
\begin{equation*}
B=\left[
\begin{array}{ccccc}
0  & -a(u_{2},u_{1})  & \dots & -a(v_{2},u_{1})  & a(u_{1},v_{1})\\
a(u_{2},u_{1})  & 0 & \dots & a(u_{2},v_{2})  & a(u_{2},v_{1})\\
\vdots&&\ \ddots && \dots \\
a(v_{2},u_{1})  & -a(u_{2},v_{2}) & \dots & 0  & a(v_{2},v_{1})\\
 -a(u_{1},v_{1}) & -a(u_{2},v_{1})  & \dots & -a(v_{2},v_{1})  & 0\\
\end{array} \right],
\end{equation*}
This shows that $B=\Psi_{id}$.

Now, we want to show that the statement holds for any $\pi \in S_n$.
We have just shown that
\begin{equation*}
\langle \psi(u_{n})\psi(v_{n}) \cdots \psi(u_{1})\psi(v_{1})
v_{\emptyset},v_{\emptyset} \rangle=\Pf(B(u_n,v_n, \dots, u_1,v_1)),
\end{equation*}
where $B$ is given above, i.e.
\begin{equation*}
B(i,j)=
\begin{cases}
B((u_i,i),(u_j,j)) & \text{ $1 \leq i<j \leq n$,}\\
B((u_i,i),(v_j',j')) & \text{ $1 \leq i \leq n < j \leq 2n$,}\\
B((v_i',i'),(v_j',j')) & \text{ $n<i < j \leq 2n$,}
\end{cases}
\end{equation*}
where $i'=2n-i+1,j'=2n-j+1$ and
\begin{equation*}
B((z,i),(w,j))=
\begin{cases}
\langle \psi(z)\psi(w)v_{\emptyset},v_{\emptyset} \rangle & \text{ $i \geq j$,}\\
-\langle \psi(w)\psi(z)v_{\emptyset},v_{\emptyset} \rangle & \text{
$i<j$}.
\end{cases}
\end{equation*}
Then \begin{equation*}
\langle \psi(u_{\pi(n)})\psi(v_{\pi(n)})
\cdots \psi(u_{\pi(1)})\psi(v_{\pi(1)}) v_{\emptyset},v_{\emptyset}
\rangle=\Pf(B(u_{\pi(n)},v_{\pi(n)}, \dots, u_{\pi(1)},v_{\pi(1)})).
\end{equation*}
Change the order of rows and columns in $B(u_{\pi(n)},v_{\pi(n)},
\dots, u_{\pi(1)},v_{\pi(1)})$ in such a way that the rows and
columns of the new matrix appear ``in order'' $u_1,u_2, \dots,
u_n,v_n, \dots ,v_2,v_1$. Let $C$ be that new matrix. The number of
switches we are making is even because we can first change the order
to $u_n,v_n,\dots,u_1,v_1$ by permuting pairs $(u_jv_j)$, and the
number of switches from this order to $u_1,u_2, \dots, u_n,v_n,
\dots ,v_2,v_1$ is $n(n-1)$ as noted above. Thus $\Pf(C)=\Pf(B)$.
Then $C=\Psi_\pi$ because
\begin{equation*}
C(i,j)=
\begin{cases}
B((u_i,\pi^{-1}(i)),(u_j,\pi^{-1}(j))) & \text{ $1 \leq i<j \leq n$,}\\
B((u_i,\pi^{-1}(i)),(v_j',\pi^{-1}(j'))) & \text{ $1 \leq i \leq n < j \leq 2n$,}\\
B((v_i',\pi^{-1}(i')),(v_j',\pi^{-1}(j'))) & \text{ $n <i< j \leq
2n$}.
\end{cases}
\end{equation*}
\end{proof}

\begin{lemma} \lb{PfaffianPsi2}
Let $\pi \in S_n$. In the domain $|u_{\pi(n)}|>|v_{\pi(n)}|> \cdots >|u_{\pi(1)}|>|v_{\pi(1)}|$
\begin{equation*}
\langle \prod_{i=1}^n
\psi(u_{\pi(i)})\psi(v_{\pi(j)})v_{\emptyset},v_{\emptyset} \rangle=\Pf(\Psi),
\end{equation*}
where $\Psi$ is a skew-symmetric $2n$-matrix given with
\begin{equation*}
\Psi(i,j)=
\begin{cases} \psi(u_i,u_j) & \text{ $1 \leq i<j \leq n$,}\\
\psi(u_i,v_j') & \text{ $1 \leq i \leq n < j \leq 2n$,}\\
\psi(v_i',v_j') & \text{ $n< i < j \leq 2n$,}
\end{cases}
\end{equation*}
where $i'=2n-i+1,j'=2n-j+1$ and
\begin{equation*}
\psi(z,w)=\frac{z-w}{4(z+w)}.
\end{equation*}
\end{lemma}
\begin{proof}
This is a direct corollary of Lemma \ref{PfaffianPsi1} and a formula given in Appendix:
\begin{equation*}
\langle \psi(z)\psi(w)v_{\emptyset},v_{\emptyset} \rangle =
\frac{z-w}{4(z+w)}\;\;\; {\rm for} \; |z|>|w|.
\end{equation*}

It is enough to check for $1 \leq i < j \leq n$, because other cases
are similar. If $\pi^{-1}(i)>\pi^{-1}(j)$, respectively
$\pi^{-1}(i)<\pi^{-1}(j)$ then
\begin{equation} \lb{same}
\psi_{\pi}((u_i,i),(u_j,j))=\psi(u_i,u_j).
\end{equation}
in the region $|u_i|>|u_j|$, respectively $|u_i|<|u_j|$. This means
that (\ref{same}) holds for the given region
$|u_{\pi(n)}|>|v_{\pi(n)}|> \cdots
>|u_{\pi(1)}|>|v_{\pi(1)}|$.
\end{proof}
\begin{remark}
Equation (\ref{lrlemma}) appears in \cite{Mat}. Lemmas \ref{PfaffianPsi1}  and \ref{PfaffianPsi2} are not of interest in \cite{Mat} for $\pi \neq id$ because in the case of the shifted Schur measure one does not need to consider the time order.
\end{remark}

\bigskip

\section{Measure on strict plane partitions} \lb{s3}

In this section we introduce strict plane partitions and a measure
on them. This measure can be obtained as a special case of the
shifted Schur process by a suitable choice of specializations of the
algebra of symmetric functions. Then the correlation functions for
this measure can be obtained as a corollary of Theorem \ref{CorFun}.
Using this result we compute the (bulk) asymptotics of the
correlation functions as the partitions become large. One of the
results we get along the way is an analog of MacMahon's formula for
the strict plane partitions.

\subsection{Strict plane partitions} \lb{s3.1}

\begin{figure} []
\centering \includegraphics[height=5cm]{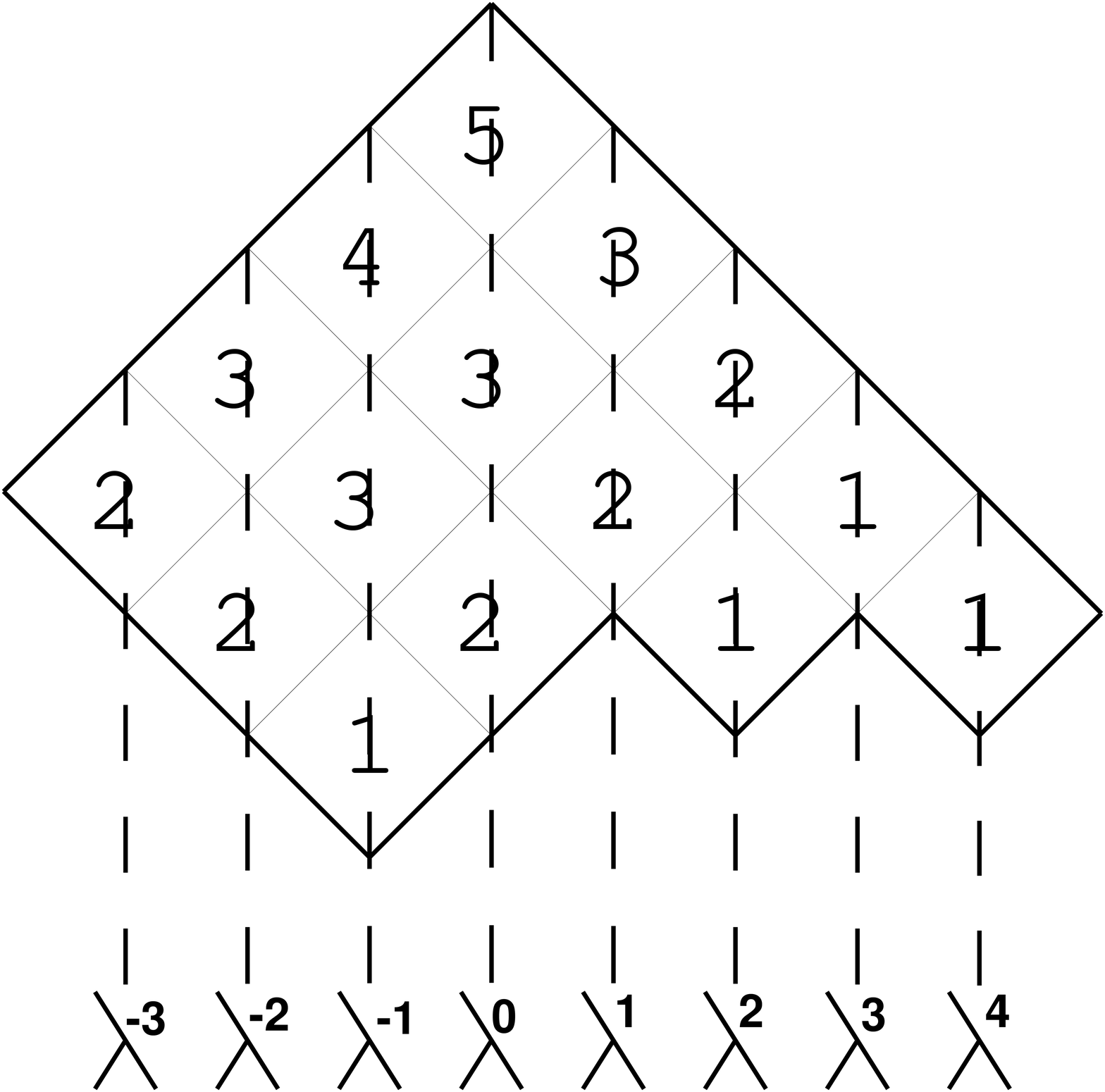}
\caption{ A strict plane partition
$\pi=((2),(3,2),(4,3,1),(5,3,2),(3,2),(2,1),(1),(1))$}
\label{PlanePartition} \end{figure}

A plane partition $\pi$ can be viewed in different ways. One way is
to fix a Young diagram, the support of the plane partition, and then
to associate a positive integer to each box in the diagram such that
integers form nonincreasing rows and columns. Thus, a plane
partition is a diagram with row and column nonincreasing integers associated to its boxes.
It can also be viewed as a finite two-sided sequence of ordinary
partitions, since each diagonal in the support diagram represents a
partition. We write $ \pi=(\lambda^{-T_L}, \ldots,
\lambda^{-1},\lambda^{0},\lambda^{1}, \ldots ,\lambda^{T_R}),$ where
the partition $\lambda^{0}$ corresponds to the main diagonal and
$\lambda^k$ corresponds to the diagonal that is shifted by $\pm k$,
see Figure \ref{PlanePartition}. Every such two-sided sequence of
partitions represents a plane partition if and only if
\begin{equation}\lb{condpp}
\begin{array}{c}
\lambda^{-T_L} \subset \cdots \subset \lambda^{-1} \subset
\lambda^{0} \supset \lambda^{1} \supset \cdots \supset
\lambda^{T_R},
\medskip
\\
\text{skew diagrams $\lambda^{k+1}-\lambda^k$
$(\lambda^{k}-\lambda^{k+1})$ are horizontal strips.}
\end{array}
\end{equation}

A strict plane partition is a finite two-sided sequence of strict
partitions with property (\ref{condpp}). In other words, it is a
plane partition whose diagonals are strict partitions. As an
example, a strict plane partition corresponding to
$\pi=((2),(3,2),(4,3,1),(5,3,2),(3,2),(2,1),(1),(1))$ is given in
Figure \ref{PlanePartition}.

\begin{figure} [htp!]
\centering
\includegraphics[height=5cm]{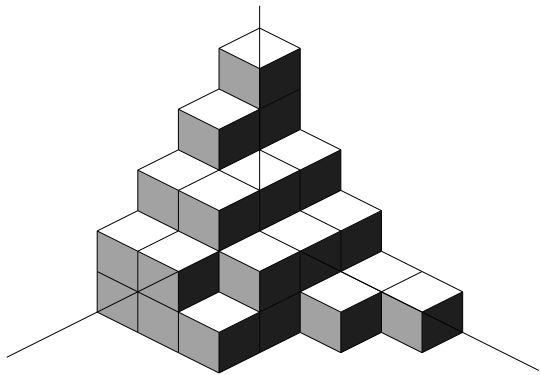}
\caption{The 3-dimensional diagram corresponding to
$\pi=((2),(3,2),(4,3,1),(5,3,2),(3,2),(2,1),(1),(1))$}
\label{3DDiagram}
\end{figure}

We give one more representation of a plane partition as a
3-dimensional diagram that is defined as a collection of $1 \times 1
\times 1$ boxes packed in a 3-dimensional corner in such a way that
the column heights are given by the filling numbers of the plane
partition. A 3-dimensional diagram corresponding to the example
above is shown in Figure \ref{3DDiagram}.

The number $|\pi|$ is the norm of $\pi$ and is equal to the sum of
the filling numbers. If $\pi$ is seen as a 3-dimensional diagram
then $|\pi|$ is its volume.

We introduce a number $A(\pi)$ that we call the alternation of $\pi$
as
\begin{equation} \lb{alternation}
A(\pi)=\sum_{i=1}^{T_R+1}a(\lambda^{i-1}-\lambda^{i})+\sum_{i=0}^{-T_L}a(\lambda^{i}-\lambda^{i-1})-l(\lambda^{0}),
\end{equation}
where $\lambda^{T_R+1}=\lambda^{-T_L-1}=\emptyset$, and $l$ and $a$
are as defined in Appendix, namely $l(\lambda)$ is the number of
(nonzero) parts of $\lambda$ and $a(\lambda-\mu)$ is the number of
connected components of the shifted skew diagram $\lambda-\mu$.

\begin{figure} [htp!]
\centering
\includegraphics[height=4cm]{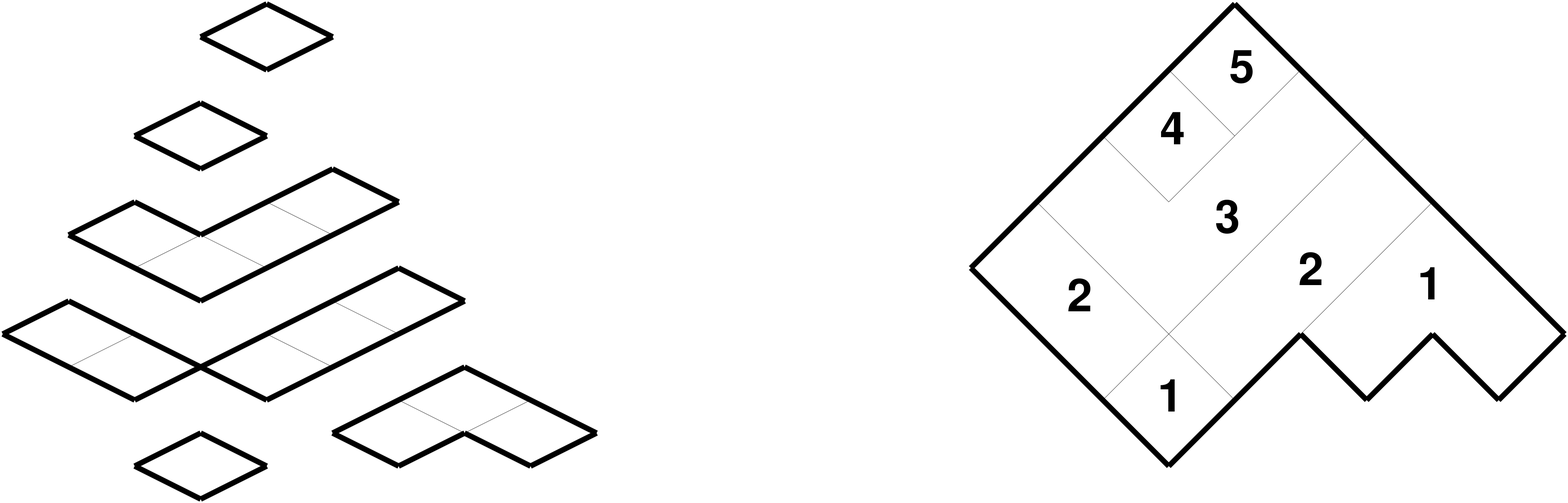}
\caption{The connected components of
$\pi=((2),(3,2),(4,3,1),(5,3,2),(3,2),(2,1),(1),(1))$}
\label{Alternate}
\end{figure}

This number is equal to the number of white islands (formed by white
rhombi) of the 3-dimensional diagram of the strict plane partition.
For the given example $A(\pi)$ is 7 (see Figure \ref{Alternate}). In
other words, this number is equal to the number of connected
components of a plane partition, where a connected component
consists of boxes associated to a same integer that are connected by
sides of the boxes (i.e. it is a border strip filled with a same
integer). For the given example there are two connected components
associated to the number 2 (see Figure \ref{Alternate}).

It is not obvious that $A(\pi)$ defined by (\ref{alternation}) is
equal to the number of the connected components. We state this fact
as a proposition and prove it.
\begin{proposition}\lb{A}
Let $\pi$ be a strict plane partition. Then the alternate $A(\pi)$
defined with (\ref{alternation}) is equal to the number of connected
components of $\pi$.
\end{proposition}
\begin{proof}
We show this inductively. Denote the last nonzero part in the last
row of the support of $\pi$ by $x$. Denote a new plane partition
obtained by removing the box containing $x$ with $\pi'$.

We want to show that $A(\pi)$ and $A(\pi')$ satisfy the relation
that does not depend whether we choose (\ref{alternation}) or the
number of connected components for the definition of the alternate.

We divide the problem in four cases I, II, III and IV shown in
Figure \ref{ProofAlternate} and we further divide these cases in several new ones. The cases
depend on the position and the value of $x_L$ and $x_R$.
\begin{figure} [htp!] \centering
\includegraphics[height=10cm]{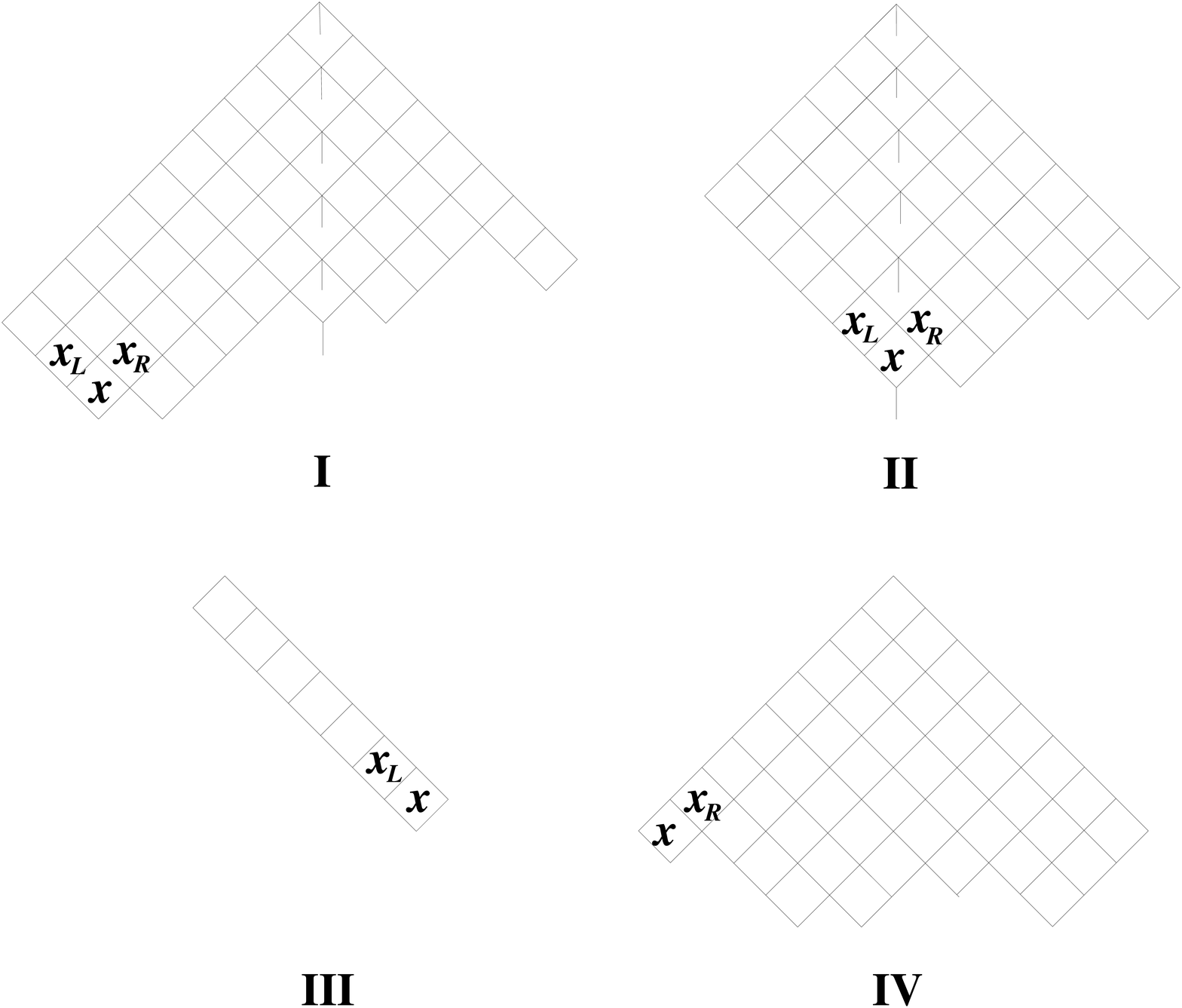}
\caption{Cases I, II, III and IV} \label{ProofAlternate}
\end{figure}

Then using (\ref{alternation}) we get
$$
A(\pi')=A(\pi)+{\text {contribution from }} x_L+{\text {contribution
from }} x_R +{\text {change of }} l(\lambda^{0}).
$$

Let us explain this formula for case I when $x_L=x_R=x$ in more detail.

Let $\lambda_{x_L}$, $\lambda_{x}$ and $\lambda_{x_R}$ be the diagonal partitions of $\pi$
containing $x_L$, $x$ and $x_R$, respectively. Let $\lambda'_{x}$ be a partition obtained from $\lambda_x$ by removing $x$. Then
\begin{eqnarray*}
{\text {contribution from }x_L}&=&a(\lambda'_x-\lambda_{x_L})-a(\lambda_x-\lambda_{x_L})=0,\\
{\text {contribution from }x_R}&=&a(\lambda_{x_R}-\lambda'_x)-a(\lambda_{x_R}-\lambda_{x})=1,\\
{\text {change of
}l(\lambda^0)}&=&l(\lambda^{0}(\pi))-l(\lambda^{0}(\pi'))=0.
\end{eqnarray*}
For all the cases the numbers are
$$A(\pi')=A(\pi)+
\begin{tabular}{|c|c|c|c|c|}
\hline
{\rm I}&{\rm II}&{\rm III}&{\rm IV}&\\
  \hline
   0+1+0 &0+0+1 &0+$\emptyset$\text{+0}&$\emptyset$\text{+0+0}& $x_L=x,x_R=x$\\
  \hline
     -1+0+0 &-1-1+1 &\-1+$\emptyset$\text{+0}&$\emptyset$\text{-1+0}&$x_L>x,x_R>x$\\
  \hline
     0+0+0 &0-1+1 &&&$x_L=x,x_R>x$\\
  \hline
     -1+1+0 &-1+0+1 &&&$x_L>x,x_R=x$\\
  \hline
\end{tabular}
$$

It is easy to verify that we get the same value for $A(\pi')$ in
terms of $A(\pi)$ using the connected component definition. Then
inductively this gives a proof that the two definitions are the
same.
\end{proof}

\begin{figure} [htp!]
\centering
\includegraphics[height=4cm]{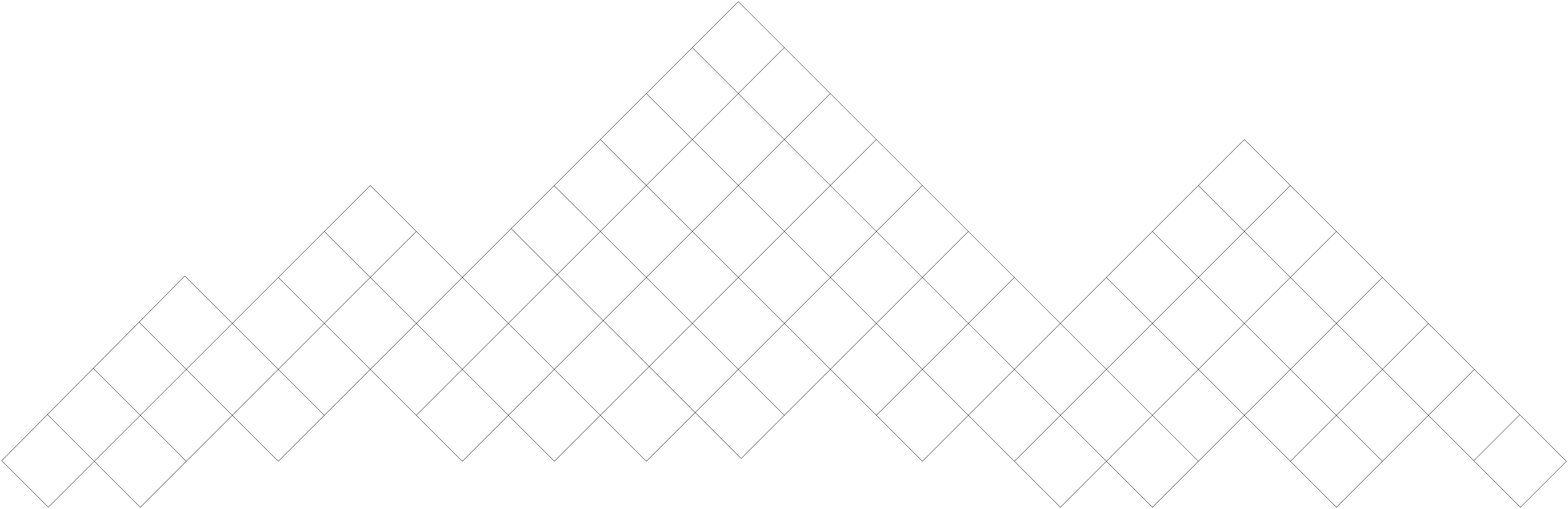}
\caption{SSPP} \label{SSPP}
\end{figure}

%
%
%
We can give a generalization of Lemma \ref{A}. Instead of starting
from a diagram of a partition we can start from a type of a diagram
shown in Figure \ref{SSPP} (connected skew Young diagram) and fill
the boxes of this diagram with row and column nonincreasing integers
such that integers on the same diagonal are distinct.
Call this object a skew strict plane partition (SSPP). We define
connected components of SSPP in the same way as for the strict plane
partitions. Then we can give an analog of Lemma \ref{A} for SSPP. We
will not use this result further in the paper.
\begin{lemma}\lb{GeneralA}
Let $\pi$ be a SSPP
$$
\pi=(\cdots \subset \lambda^{t_{-2}} \supset \cdots \supset
\lambda^{t_{-1}}\subset \cdots \subset \lambda^{t_0} \supset \cdots
\supset \lambda^{t_1}\subset \cdots \subset \lambda^{t_2}\supset
\cdots).
$$
Then the alternate $A(\pi)$ defined by
$$
A(\pi)=\sum_{i=-\infty}^{\infty}a(\lambda^i,\lambda^{i-1})-
\sum_{i=-\infty}^{\infty}l(\lambda^{t_{2i}})+
\sum_{i=-\infty}^{\infty}l(\lambda^{t_{2i+1}}),
$$
with
$$
a(\lambda,\mu)=
\begin{cases}
a(\lambda-\mu) &
\lambda \supset \mu\\
a(\mu-\lambda) & \mu \supset \lambda
\end{cases}
$$
is equal to the number of connected components of $\pi$.
\end{lemma}
\begin{proof}
The proof is similar to that of Lemma \ref{A}.
\end{proof}

\subsection{The measure} \lb{s3.2}
For $q$ such that $0<q<1$ we define a measure on strict plane
partitions by
\begin{equation}\lb{measure}
P(\pi)=\frac{2^{A(\pi)}q^{|\pi|}}{Z},
\end{equation}
where $Z$ is the partition function, i.e $Z=\sum_\pi
2^{A(\pi)}q^{|\pi|}$. This measure has a natural interpretation since
$2^{A(\pi)}$ is equal to the number of colorings of the connected
components of $\pi$.

This measure can be obtained as a special shifted Schur process for
an appropriate choice of the specializations $\rho$. Then
Proposition \ref{Z} can be used to obtain the partition function:
\begin{proposition} \lb{SMM}
$$
\sum_{\substack {\pi \text{ is a strict}\\ \text{plane partition}}
} 2^{A(\pi)}q^{|\pi|}=\prod_{n=1}^{\infty}\left(
\frac{1+q^n}{1-q^n}\right)^n.
$$
\end{proposition}
This is the generating formula for the strict plane partitions and we call it the shifted
MacMahon's formula since it can be viewed as an analog of  MacMahon's generating formula for the plane partitions that says
$$
\sum_{\substack {\pi \text{ is a plane}\\ \text{partition}} }
q^{|\pi|}=\prod_{n=1}^{\infty}\left( \frac{1}{1-q^n}\right)^n.
$$

Before we show that this measure is a special shifted Schur process
let us recall some facts that can be found in Chapter 3 of
\cite{Mac}. We need the values of skew Schur $P$ and $Q$ functions
for some specializations of the algebra of symmetric functions. They
can also be computed directly using (\ref{Qtableau}).

If $\rho$ is a specialization of $\Lambda$ where
$x_1=s,\,x_2=x_3=\ldots=0$ then
$$
\begin{array}{lcc}
 Q_{\lambda/\mu}(\rho)=
\begin{cases}
2^{a(\lambda-\mu)}s^{|\lambda|-|\mu|} & \text{$\;\;\;\;\;\;\;\;\;\;\;\;\;\;\lambda \supset \mu$,  $\lambda-\mu$ is a horizontal strip},\\
0 & \text{$\;\;\;\;\;\;\;\;\;\;\;\;\;\;$otherwise},
\end{cases}\\
P_{\lambda/\mu}(\rho)=
\begin{cases}
2^{a(\lambda-\mu)-l(\lambda)+l(\mu)}s^{|\lambda|-|\mu|} & \text{ $\lambda \supset \mu$, $\lambda-\mu$ is a horizontal strip},\\
0 & \text{ otherwise},
\end{cases}
\end{array}$$
where, as before, $a(\lambda-\mu)$ is the number of connected
components of the shifted skew diagram $\lambda-\mu$. In particular,
if $\rho$ is a specialization given with $x_1=x_2=\ldots=0$ then
$$
\begin{array}{ccc}
 Q_{\lambda/\mu}(\rho)=
\begin{cases}
1 & \text{ $\lambda = \mu$,}\\
0 & \text{ otherwise},
\end{cases}\\
P_{\lambda/\mu}(\rho)=
\begin{cases}
1 & \text{ $\lambda = \mu$,}\\
0 & \text{ otherwise}.
\end{cases}
\end{array}$$

In order to obtain the measure (\ref{measure}) as a special shifted
Schur process we set
\begin{equation}{\lb{ros}}
\begin{array}{llll}
\rho_n^+:x_1=q^{-(2n+1)/2},\,x_2=x_3=\ldots=0 &&&n \leq -1,\\
\rho_n^-:x_1=x_2=\ldots=0 &&&n \leq -1,\\
\rho_n^-:x_1=q^{(2n+1)/2},\,x_2=x_3=\ldots=0 &&&n \geq 0,\\
\rho_n^+:x_1=x_2=\ldots=0 &&&n \geq 0.
\end{array}
\end{equation}

This measure is supported on strict plane partitions viewed as a two-sided sequence $(\dots,\lambda^{-n},\dots,\lambda^0,\dots,\lambda^n,\dots)$. Indeed, for any two sequences $\lambda=(\dots,\lambda^{-n},\dots,\lambda^0,\dots,\lambda^n,\dots)$ and $\mu=(\dots,\mu^{-n},\dots,\mu^0,\dots,\mu^{n},\dots)$ we have that the
weight is given with
\begin{eqnarray*} W(\lambda, \mu)&=&\prod_{n=-\infty}^{\infty} Q_{\lambda^{n}/\mu^{n-1}}
(\rho_{n-1}^{+})P_{\lambda^{n}/\mu^{n}} ({\rho_n^{-}}),
\end{eqnarray*}
where only finitely many terms contribute. Then $W(\lambda,\mu)=0$
unless
$$
\mu^{n}=\begin{cases}
\lambda^n&n<0,\\
\lambda^{n+1}&n\geq0,
\end{cases}
$$
$$
\cdots\subset\lambda^{-n}\subset \cdots\subset \lambda^0 \supset \cdots \supset \lambda^n\supset\cdots,
$$
$$
\begin{array}{c}
\text{skew diagrams } \lambda^{n+1} - \lambda ^{n} \text{ for } n< 0 \text{ and}\\
\lambda^{n} - \lambda ^{n+1} \text{ for } n\geq 0 \text{ are
horizontal strips},
\end{array}
$$
and in that case
\begin{eqnarray*} W(\lambda, \mu)&=&\prod_{n=-\infty}^{0}2^{a(\lambda^{n}-\lambda^{n-1})}q^{(-2n+1)(|\lambda^{n}|-|\lambda^{n-1}|)/2}\\ &&\cdot\prod_{n=1}^{\infty}2^{a(\lambda^{n-1}-\lambda^{n})-l(\lambda^{n-1})+l(\lambda^{n})}q^{(2n-1)(|\lambda^{n-1}|-|\lambda^{n}|)/2}\\
&=&2^{A(\lambda)}q^{|\lambda|}.
\end{eqnarray*}
Thus, the given choice of $\rho$'s define a shifted Schur process (or a limit of shifted Schur processes as we explain in a remark below) that is indeed equal to the measure on the strict plane partitions given with (\ref{measure}).

Proposition \ref{Z} allows us to obtain the shifted MacMahon's formula. If $\rho^+$ is $x_1=s,\,x_2=x_3=\ldots=0$ and $\rho^-$ is
$x_1=t,\,x_2=x_3=\ldots=0$ then
$$
H(\rho^+,\rho^-)=\prod_{i,\,j} \left. \frac{1+x_iy_j}{1-x_iy_j} \right| _{x=\rho^+,\,
y=\rho^-}=\frac{1+st}{1-st}.
$$
Thus, for the given specializations of $\rho_i^+$'s and $\rho_i^-$'s
we have
\begin{eqnarray*}
Z(\rho)=\prod_{i<j}H(\rho_i^+,\rho_j^-)&=&\frac{1+q}{1-q} \cdot \frac{1+q^2}{1-q^2} \cdot \frac{1+q^3}{1-q^3}\cdots\\
&&\frac{1+q^2}{1-q^2} \cdot \frac{1+q^3}{1-q^3}\cdots\\
&&\frac{1+q^3}{1-q^3} \cdot \frac{1+q^4}{1-q^4}\cdots\\
&=&\prod_{n=1}^{\infty}\left( \frac{1+q^n}{1-q^n}\right)^n.
\end{eqnarray*}

\begin{remark}
A shifted Schur process depends on finitely many specializations.
For that reason, measure (\ref{measure}) is a limit of measures
defined as shifted Schur processes rather than a shifted Schur
process itself. For every $T$ let specializations $\rho^{\pm}_n$ be
as in (\ref{ros}) if $|n|\leq T$ and zero otherwise. They define a
shifted Schur process whose support is $S_T$ that is the set of
strict plane partitions with $\lambda^n=\emptyset$ for every
$|n|>T$. The partition function for this measure is  $\sum_{\pi \in
S_T}2^{A(\pi)}q^{|\pi|}$ and is bounded by
$\prod(\frac{1+q^n}{1-q^n})^n$. Let $S$ be the set of all strict
plane partitions. Then the $T$th partial sum (sum of all terms that
involve $q^m$ for $m \leq T$) of $\sum_{\pi \in
S}2^{A(\pi)}q^{|\pi|}$, which is equal to the $T$th partial sum of
$\sum_{\pi \in S_T}2^{A(\pi)}q^{|\pi|}$, is bounded by
$\prod(\frac{1+q^n}{1-q^n})^n$. Hence, $\sum_{\pi \in
S}2^{A(\pi)}q^{|\pi|}$ converges.  Therefore, $\sum_{\pi \in
S_T}2^{A(\pi)}q^{|\pi|} \to \sum_{\pi \in S}2^{A(\pi)}q^{|\pi|}$ as
$T \to \infty$. Thus, the correlation function of the measure
(\ref{measure}) is the limit of the correlation functions of the
approximating shifted Schur processes as $T\to \infty$.
\end{remark}
Our next goal is to find the correlation function for the measure
(\ref{measure}). For that we need to restate Theorem \ref{CorFun}
for the given specializations. In particular, we need to determine
$J(t,z)$. When $\rho$ is such that $x_1=s,x_2=x_3=\cdots=0$ then
(see (\ref{Q}))
$$
F(\rho;z)=\frac{1+sz}{1-sz}.
$$
Thus, for our specializations (see (\ref{J}))
\begin{equation}\lb{J3D}
J(t,z)=
\begin{cases}
\frac{\displaystyle \prod_{m \geq t} \frac
{1+q^{m+1/2}z}{1-q^{m+1/2}z}}{\displaystyle \prod_{m\geq0}
\frac {1+q^{m+1/2} z^{-1}}{1-q^{m+1/2} z^{-1}}}= \displaystyle\frac{(q^{1/2} z^{-1};q)_\infty(-q^{t+1/2}z;q)_\infty}{(-q^{1/2} z^{-1};q)_\infty(q^{t+1/2}z;q)_\infty}&t\geq0\\
\\
 \frac{\displaystyle \prod_{m \geq 0} \frac
{1+q^{m+1/2}z}{1-q^{m+1/2}z}}{\displaystyle \prod_{m \geq -t} \frac {1+q^{m+1/2} z^{-1}}{1-q^{m+1/2} z^{-1}}}= \displaystyle\frac{(-q^{1/2}z;q)_\infty(q^{-t+1/2} z^{-1};q)_\infty}{(q^{1/2}z;q)_\infty(-q^{-t+1/2} z^{-1};q)_\infty}&t<0,
\end{cases}
\end{equation}
where
$$
(z;q)_\infty=\prod_{n=0}^{\infty}(1-q^nz)
$$
is the quantum dilogarithm function.


%
%

It is convenient to represent a strict plane
partition $ \pi=(\ldots,
\lambda^{-1},\lambda^{0},\lambda^{1}, \ldots )$ as a subset of
$$
\frak{X}=\left\{(t,x)\in\bbZ \times \bbZ
\left|\right.x>0\right\},
$$
where $(t,x)$ belongs to this subset if and only if $x$ is a part of $\lambda^t$.
We call this subset the plane diagram of the strict plane partition $\pi$.

\begin{corollary} \lb{CorFun3D}
For a set $X=\{(t_i,x_i):i=1, \dots, n\}\subset \frak{X}$
representing a  plane diagram, the correlation function is
given with
\begin{equation} \lb{Pfaffian3D}
\rho(X)=\Pf(M_X^{3D})
\end{equation}
where $M_X^{3D}$ is a skew-symmetric $2n \times 2n$ matrix and
$M_X^{3D}(i,j)$ is given with
\begin{equation}\lb{kernel3D}
\begin{cases} K_{x_i,x_j}(t_i,t_j) & \text{ $1 \leq i<j \leq n$,}\\
(-1)^{x_{j'}}K_{x_i,-x_{j'}}(t_i,t_{j'}) & \text{ $1 \leq i \leq n < j \leq 2n$,}\\
(-1)^{x_{i'}+x_{j'}}K_{-x_{i'},-x_{j'}}(t_{i'},t_{j'}) & \text{ $n <
i < j \leq 2n$,}
\end{cases}
\end{equation}
where $i'=2n-i+1$ and $K((t_i,x),(t_j,y))$ is the
coefficient of $z^xw^y$ in the formal power series expansion of
\begin{equation*}
K((t_i,z),(t_j,w))=\frac{z-w}{2(z+w)}J(t_i,z)J(t_j,w)
\end{equation*}
in the region $|z|>|w|$ if $t_i \geq t_j$ and $|z|<|w|$ if $t_i <
t_j$. Here $J(t,z)$ is given with (\ref{J3D}).
\end{corollary}


\bigskip

\section{Asymptotics of large random strict plane partitions} \lb{s4}
In this section we compute the bulk limit for shifted plane diagrams
with a distribution proportional to
$2^{A(\pi)}q^{|\pi|}$, where $\pi$ is the corresponding strict plane
partition. In order to determine the correct scaling we first consider the asymptotic behavior of the volume $|\pi|$ of our random strict plane partitions.

\subsection{Asymptotics of the volume} \lb{conprob}
The scaling we are going to choose when computing the limit of the
correlation functions will be $r=-\log q$ for all directions. One reason for that lies in the fact that $r^3|\pi|$
converges in probability to a constant. Thus, the scaling assures that
the volume tends to a constant. Our argument is similar to that of Lemma 2 of \cite{OR}.
\begin{proposition}
If the probability of $\pi$ is given with (\ref{measure}) then
$$
r^3|\pi| \to \frac{7}{4}\zeta(3),\;\;\;r \to +0.
$$
The convergence is in probability.
\end{proposition}

\begin{proof}
We recall that if $E(X_n) \to c$ and $Var(X_n) \to 0$ then $X_n \to c$ in
probability.
Thus, it is enough to show that
$$
E(r^3|\pi|) \to \frac{7}{4}\zeta(3),\;\;\;r \to +0
$$
and
$$
Var(r^3|\pi|)\to 0,\;\;\;r \to +0.
$$

First, we observe that
$$
E(|\pi|)=\frac{\sum_{\pi}2^{A(\pi)}q^{|\pi|}|\pi|}{Z}=\frac{q\frac{d}{dq}Z}{Z}
$$
and
$$
Var(|\pi|)=E(|\pi|^2)-E(|\pi|)^2=\frac{\sum_{\pi}2^{A(\pi)}q^{|\pi|}|\pi|^2}{Z}-\frac{q^2(\frac{d}{dq}Z)^2}{Z^2}
=q\frac{d}{dq}E(|\pi|).
$$

Since $Z=\prod_{n\geq 1}(\frac{1+q^n}{1-q^n})^n$, we have
$$
\frac{d}{dq}{Z}=\prod_{n\geq 1}\left(\frac{1+q^n}{1-q^n}\right)^n\sum_{m\geq 1} \displaystyle {\frac{
m\left(\displaystyle \frac{1+q^m}{1-q^m}\right)^{m-1}\displaystyle \frac{2mq^{m-1}}{(1-q^m)^2}}{\left(\displaystyle \frac{1+q^m}{1-q^m}\right)^m}=Z\sum_{m\geq 1}\frac{2m^2q^{m-1}}{1-q^{2m}}}.
$$
Then
$$
E(|\pi|)=\frac{q\frac{d}{dq}Z}{Z}=\sum_{m\geq 1}\frac{2m^2q^m}{1-q^{2m}}=\sum_{m\geq1,k\geq0}
2m^2q^{m(2k+1)}=2\sum_{k\geq0}\frac{q^{2k+1}(1+q^{2k+1})}{(1-q^{2k+1})^3},
$$
because
$$
\sum_{m\geq1} m^2q^m=\frac{q(1+q)}{(1-q)^3},\;\;\;|q|<1.
$$
Now,
$$
r^3\frac{q^{2k+1}(1+q^{2k+1})}{(1-q^{2k+1})^3}
=\frac{e^{-r(2k+1)}(1+e^{-r(2k+1)})}{\left(\displaystyle
\frac{r(2k+1)+o(r(2k+1))}{r}\right)^3} \nearrow
\frac{1}{(2k+1)^3}\;\;\;r\to+0.
$$
Then by the uniform convergence
$$
\lim_{r \to +0}r^3E(|\pi|)=2\sum_{k\geq0} \lim_{r \to
+0}r^3\frac{q^{2k+1}(1+q^{2k+1})}{({1-q^{2k+1}})^3}=2\sum_{k\geq0}\frac{1}{(2k+1)^3}.
$$
Finally, since
$$
\sum_{k\geq0}\frac{1}{(2k+1)^3}=(1-\frac{1}{2^3})\zeta(3)
$$
it follows that
$$
E(r^3|\pi|) \to \frac{7}{4}\zeta(3),\;\;\;r \to+0.
$$

For the variance we have
$$
Var(|\pi|)=q\frac{d}{dq}E(|\pi|)=-\frac{d}{dr}E(|\pi|) \sim
\frac{21\zeta(3)}{4r^4},
$$
since by L'H\^opital's rule
$$
\frac{7}{4}\zeta(3)=\lim_{r \to
+0}\frac{E(|\pi|)}{\frac{1}{r^3}}=\lim_{r \to
+0}\frac{\frac{d}{dr}E(|\pi|)}{-3\frac{1}{r^4}}.
$$
Thus,
$$
Var(r^3|\pi|) \to 0,\;\;\;r \to+0.
$$
\end{proof}

\subsection{Bulk scaling limit of the correlation functions} \lb{s4.2}
We compute the limit of the correlation function for a set of points that when scaled by $r=-\log q$
tend to a fixed point in the plane as  $r \to +0$. Namely, we compute the limit of
(\ref{kernel3D}) for
$$
rt_i \to \tau, \;\;\; rx_i \to \chi \;\;\; \text{as } r \to +0,
$$
$$ t_i-t_j= \Delta t_{ij}=\text{const}, \;\;\; x_i-x_j= \Delta x_{ij}=
\text{const},
$$
where $\chi \geq 0$.

In Theorem \ref{limcorfun} we show that in the limit the Pfaffian
(\ref{Pfaffian3D}) turns into a determinant whenever
$\chi>0$ and it remains a Pfaffian on the boundary
$\chi=0.$

Throughout this paper $\gamma_{R,\theta}^+$ ($\gamma_{R,\theta}^-$)
stands for the counterclockwise (clockwise) oriented arc on $|z|=R$
from $Re^{-i\theta}$ to $Re^{i\theta}$.

More generally, if $\gamma$ is a curve parameterized by
$R(\phi)e^{i\phi}$  for $\phi \in[-\pi,\pi]$ then $\gamma^+_\theta$
$(\gamma^-_\theta)$ stands for the counterclockwise (clockwise)
oriented arc on $\gamma$ from $R(\theta)e^{-i\theta}$ to
$R(\theta)e^{i\theta}$.

Recall that our phase space is $\frak{X}=\left\{(t,x)\in \bbZ \times \bbZ
\left|\right. x>0 \right\}.$

\begin{theorem} \lb{limcorfun}
Let $X=\{(t_i,x_i):i=1, \dots, n\}\subset \frak{X}$ be such that
$$
rt_i \to \tau, \;\;\; rx_i \to \chi \;\;\; \text{as } r \to +0,
$$
$$ t_i-t_j= \Delta t_{ij}=\text{const}, \;\;\; x_i-x_j= \Delta x_{ij}=
\text{const}.
$$
a) If $\chi > 0$ then
$$
\lim_{r \to +0}\rho(X) = \det[K(i,j)]_{i,j=1}^n,
$$
where
$$
K(i,j)=\frac{1}{2\pi i} \int_{\gamma_{R,\theta} ^\pm} \left(
\frac{1-z}{1+z} \right)^{\Delta t_{ij}} \frac{1}{z^{\Delta
x_{ij}+1}}dz,
$$
where we choose $\gamma_{R,\theta} ^+$ if $\Delta t_{ij} \geq 0$ and
$\gamma_{R,\theta} ^-$ otherwise, where $R=e^{-|\tau|/2}$ and
\begin{equation}{\lb{theta}}
\theta=\begin{cases}\arccos\displaystyle\frac{(e^{|\tau|}+1)(e^{\chi}-1)}{2e^{|\tau|/2}{(e^{\chi}+1)}},
&\displaystyle\frac{(e^{|\tau|}+1)(e^{\chi}-1)}{2e^{|\tau|/2}{(e^{\chi}+1)}}\leq 1,\\
0,&\text{otherwise,}
\end{cases}
\end{equation}

b) If $\chi=0$ and in addition to the above conditions we
assume
$$
x_i=\text{const}
$$
then
$$
\lim_{r \to +0}\rho(X) = \Pf [M(i,j)]_{i,j=1}^{2n},
$$
where $M$ is a skew symmetric matrix given by
$$
M(i,j)=\begin{cases} \displaystyle \frac{(-1)^{x_j}}{2\pi i}
\int_{\gamma_{R,\theta} ^\pm}
\left(\frac{1-z}{1+z} \right)^{\Delta t_{ij}}\frac{dz}{z^{x_i+x_j+1}}& \text{ $1 \leq i<j \leq n$,}\\
 \displaystyle \frac{1}{2\pi i}
\int_{\gamma_{R,\theta} ^\pm}
\left(\frac{1-z}{1+z} \right)^{\Delta t_{ij'}}\frac{dz}{z^{x_i-x_{j'}+1}}&\text{ $1 \leq i \leq n < j \leq 2n$,}\\
\displaystyle \frac{(-1)^{x_{i'}}}{2\pi i} \int_{\gamma_{R,\theta}
^\pm} \left(\frac{1-z}{1+z} \right)^{\Delta
t_{i'j'}}\frac{dz}{z^{-(x_{i'}+x_{j'})+1}}\ & \text{ $n < i < j \leq
2n$,}
\end{cases}
$$
where $i'=2n-i+1$ and we choose $\gamma _{R,\theta} ^+$ if $\Delta
t_{ij} \geq 0$ and $\gamma _{R,\theta} ^-$ otherwise, where
$R=e^{-|\tau|/2}$ and $\theta=\pi/2$.
\end{theorem}
This implies that for an equal time configuration (points on the same vertical line) we get
\begin{corollary}\lb{SineKernel}
For $X=\{(t,x_i),i=1, \dots ,n\}\subset \frak{X}$ such that
$$
rt \to \tau, \;\;\; rx_i \to \chi \;\;\; \text{as } r \to +0,
$$
$$x_i-x_j= \Delta x_{ij}=
\text{const},
$$
where $\chi \geq 0$,
$$
\lim_{r\to +0}\rho(X)= \det \left[ \frac{\sin(\theta \Delta x_{ij})}{\pi \Delta
x_{ij}} \right],
$$
where $\theta$ is given with (\ref{theta}).
\end{corollary}
\begin{remark}
The kernel of Theorem \ref{limcorfun} can be viewed as an extension of the discrete sine kernel of Corollary \ref{SineKernel}. This is one of the extensions constructed in Section 4 of \cite{B}, but it is the first time that this extension appears in a ``physical'' problem.
\end{remark}
The limit of the 1-point correlation function gives a density for the points of the plane diagram of
strict plane partitions. We state this as a corollary.
\begin{corollary}\lb{density}
The limiting density of the point $(\tau,\chi)$ of the plane diagram of
strict plane partitions   is given with
$$
\rho(\tau,\chi)= \lim_{\substack {rt \to \tau, rx \to \chi,\\ r
\to+0}}
K_{3D}((t,x),(t,x))=\frac{\theta}{\pi},
$$
where $\theta$ is given with (\ref{theta}).
\end{corollary}
\begin{remark}
Using Corollary \ref{density} we can determine the hypothetical limit shape of a
typical 3-dimensional diagram. This comes from the
observation that for a strict plane partition
\begin{eqnarray*}
x(\tau,\chi)&=&\int_{\chi}^{\infty}\rho(\tau,s)ds,\\
y(\tau,\chi)&=&\int_{\chi}^{\infty}\rho(\tau,s)ds+\tau,\\
z(\tau,\chi)&=&\chi,
\end{eqnarray*}
if $\tau \geq 0$ and
\begin{eqnarray*}
x(\tau,\chi)&=&\int_{\chi}^{\infty}\rho(\tau,s)ds-\tau,\\
y(\tau,\chi)&=&\int_{\chi}^{\infty}\rho(\tau,s)ds,\\
z(\tau,\chi)&=&\chi
\end{eqnarray*}
if $\tau<0$. 
Indeed, for a point $(x,y,z)$ of a 3-dimensional diagram, where $z=z(x,y)$ is the height of the column based at $(x,y)$, the corresponding point in the plane diagram is $(\tau,\chi)=(y-x,z)$. Also, for the right (respectively left) part of a 3-dimensional diagram the coordinate $x$ (respectively $y$) is given by the number of points of the plane diagram above $(\tau,\chi)$ (i.e. all $(\tau,s)$ with $s\geq \chi$) and that is in the limit equal to $\int_{\chi}^{\infty}\rho(\tau,s)ds$.

The hypothetical limit shape is shown in Figure \ref{limitshapefig}.


\begin{figure}[htp!]
\subfigure[The limit shape]
{\includegraphics[height=5cm]{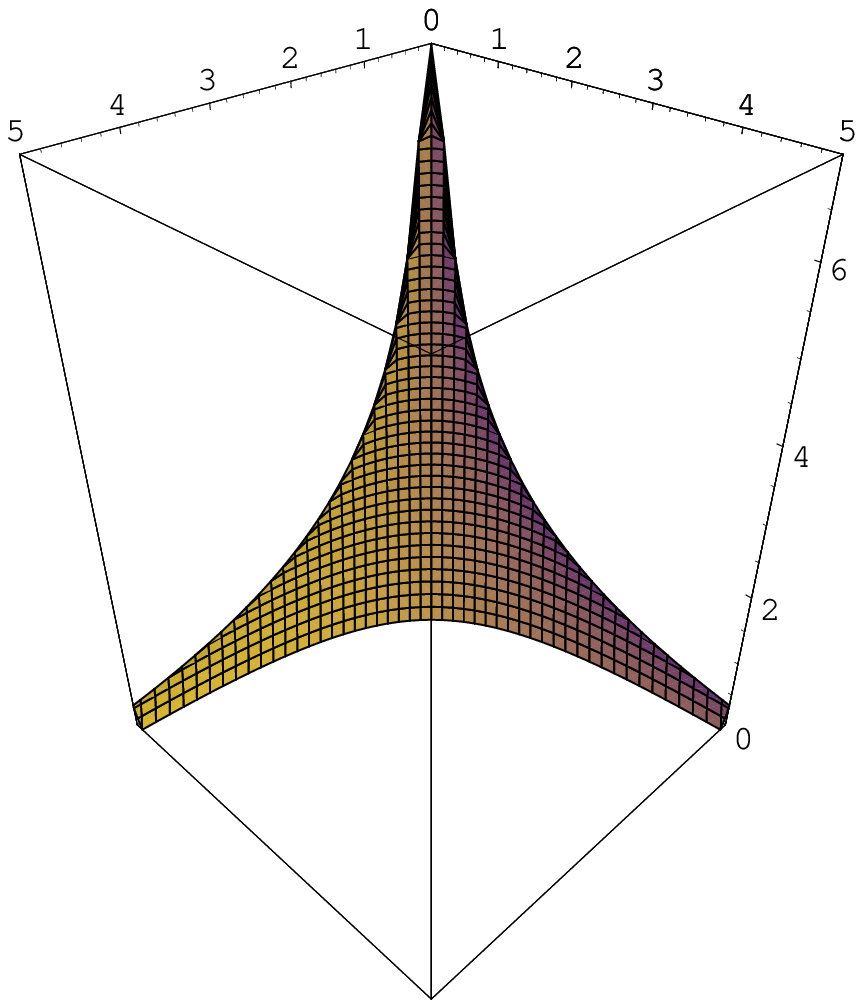}\lb{limitshapefig}}
\hspace{1cm}
\subfigure[The amoeba of -1+z+w+zw]
{\includegraphics[height=5cm]{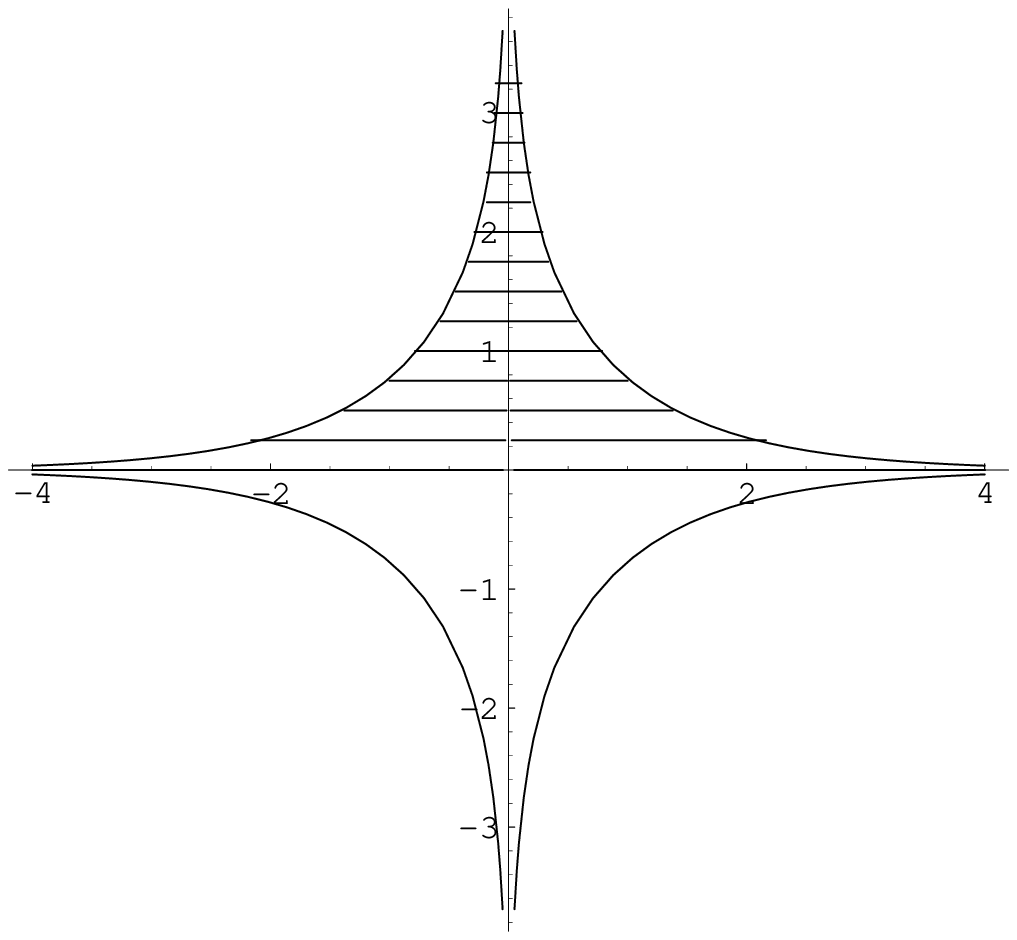}\lb{amoebafig}}
\caption{}
\end{figure}

Recall that the amoeba of a polynomial $P(z,w)$ where
$(z,w)\in\bbC^2$ is
$$
\{(\xi,\omega)=(\log|z|,\log|w|)\in
\bbR^2\;|\;(z,w)\in(\bbC\backslash \{0\})^2, \; P(z,w)=0\}.
$$
The limit shape of the shifted Schur process can be parameterized
with $(\xi, \omega)=(\tau/2,\chi/2)$ for $(\xi, \omega) \in
\mathfrak{D}$ where $\mathfrak{D}$ is the shaded region given in
Figure \ref{amoebafig}. The boundaries of this region are $\omega=0$,
$\omega=\log[(e^\xi+1)/(e^\xi-1)]$ for $\xi >0$ and
$\omega=\log[(e^{-\xi}+1)/(e^{-\xi}-1)]$ for $\xi <0$. This region
is the half of the amoeba of $-1+z+w+zw$ for $\omega=\log|w|\geq0$.
\end{remark}

\begin{proof} (Theorem \ref{limcorfun})
Because of the symmetry it is enough to consider only $\tau \geq 0$.

In order to compute the limit of (\ref{kernel3D}) we need to
consider three cases
\begin{eqnarray*}
1)\; K_{++}^{1,2}&=&K_{x_1,x_2}(t_1,t_2)\\
2)\; K_{+-}^{1,2}&=&(-1)^{x_2}K_{x_1,-x_2}(t_1,t_2)\\
3)\; K_{--}^{1,2}&=&(-1)^{x_1+x_2}K_{-x_1,-x_2}(t_1,t_2),
\end{eqnarray*}
when $rx_i \to \chi, \, rt_i \to \tau, \, r \to +0, \, q=e^{-r}$
and $\Delta t_{1,2}$ and $\Delta x_{1,2}$ are fixed.\\

Since we are interested in the limits of 1), 2) and 3) we can assume
that $rt_i \geq e^{\tau/2}$, for every $i$.

We start with 2).  By the definition,
$$
K_{+-}^{1,2}=\frac{(-1)^{x_2}}{(2\pi i)^2} \iint
\limits_{\substack{|z|=(1 \pm \epsilon){e^{\tau/2}} \\{|w|=(1 \mp
\epsilon)e^{\tau/2}}}} \frac {z-w}{2(z+w)}J(t_1,z)J(t_2,w)
\frac{1}{z^{x_1+1}w^{-x_2+1}}dzdw,
$$
where we take the upper signs if
$t_1 \geq t_2$ and the lower signs otherwise. Here, we choose $\epsilon \in (0,1-{q}^{1/2})$ since $J(t,z)$ is equal to its power series expansion in the region $q^{1/2}<|z|<q^{-t-1/2}$. With the change of variables $w
\mapsto -w$ we get
$$
K_{+-}^{1,2}=e^{-\tau (x_1-x_2)/2}\cdot I_{+-},
$$
where
$$
I_{+-}=\frac{1}{(2\pi i)^2} \iint \limits_{\substack{|z|=(1 \pm
\epsilon){e^{\tau/2}} \\{|w|=(1 \mp \epsilon)e^{\tau/2}}}}  \frac
{z+w}{2zw(z-w)} \frac{J(t_1,z)}{J(t_2,w)} \left(
\frac{z}{e^{\tau/2}} \right)^{-x_1}
\left(\frac{w}{e^{\tau/2}}\right)^{x_2}dzdw.
$$

We consider cases 1) and 3) together
because terms in the Pfaffian contain $K_{++}$ and $ K_{--}$ in pairs.

Using the definition and a simple change of coordinates $w \mapsto
-w$ we have that
\begin{equation}
K_{++}^{1,2} \cdot
K_{--}^{3,4}=e^{-\tau(x_1+x_2-x_3-x_4)/2}(-1)^{x_2+x_3}\cdot I_{++}
\cdot I_{--} \lb{constI++I--}
\end{equation}
where
$$
I_{++}=\frac{1}{(2\pi i)^2} \iint \limits_{\substack{|z|=(1 \pm
\epsilon){e^{\tau/2}} \\{|w|=(1 \mp \epsilon)e^{\tau/2}}}}  \frac
{z+w}{2zw(z-w)} \frac{J(t_1,z)}{J(t_2,w)} \left(
\frac{z}{e^{\tau/2}} \right)^{-x_1}
\left(\frac{w}{e^{\tau/2}}\right)^{-x_2}dzdw
$$
and
$$
I_{--}=\frac{1}{(2\pi i)^2} \iint \limits_{\substack{|z|=(1 \pm
\epsilon){e^{\tau/2}} \\{|w|=(1 \mp \epsilon)e^{\tau/2}}}}  \frac
{z+w}{2zw(z-w)} \frac{J(t_3,z)}{J(t_4,w)} \left(
\frac{z}{e^{\tau/2}} \right)^{x_3}
\left(\frac{w}{e^{\tau/2}}\right)^{x_4}dzdw,
$$
where the choice of signs is as before: in the first integral we choose the upper signs if $t_1 \geq t_2$ and the lower signs otherwise; similarly in the second integral with $t_3$ and $t_4$.

In the first step we will prove the following claims.
\begin{claim1}
$$\lim_{r\to +0} I_{+-}=\frac{1}{2\pi
i}\int_{\gamma_{e^{\tau/2},
\theta}^\pm}\frac{1}{z}\left(\frac{1-e^{-\tau}z}{1+e^{-\tau}z}\right)^{\Delta
t_{1,2}}\left( \frac{z}{e^{\tau/2}}\right)^{-\Delta x_{1,2}}dz,
$$
where we choose the plus sign if $t_1 \geq t_2$ and minus otherwise
and $\theta$ is given with (\ref{theta}).
\end{claim1}

\begin{claim2}
$$\lim_{r\to +0} I_{++}=\lim_{r\to +0} \frac{1}{2\pi
i}(-1)^{x_1+x_2+1}\int_{\gamma_{e^{\tau/2},\theta}^\mp}\frac{1}{z}\frac{J(t_2,z)}{J(t_1,z)}\left(
\frac{z}{e^{\tau/2}}\right)^{-(x_1+x_2)}dz,
$$
where we choose the minus sign if $t_1 \geq t_2$ and plus otherwise,
where $\theta$ is given with (\ref{theta}). (We later show that the
limit in the right hand side always exists.)
\end{claim2}

\begin{claim3}
$$\lim_{r\to +0} I_{--}=\lim_{r\to +0} \frac{1}{2\pi
i}\int_{\gamma_{e^{\tau/2},\theta}^
\pm}\frac{1}{z}\frac{J(t_3,z)}{J(t_4,z)}\left(
\frac{z}{e^{\tau/2}}\right)^{x_3+x_4}dz,
$$
where we choose the plus sign if $t_3 \geq t_4$ and minus otherwise,
where $\theta$ is given with (\ref{theta}). (We later show that the
limit in the right hand side always exists.)

\end{claim3}

In the next step we will show that Claims 2 and 3 imply that the limits of $I_{++}$ and $I_{--}$
vanish when $r \to +0$ unless $\chi=0$. We state this in two
claims.

\begin{claim4}
If $\chi>0$ then $\lim_{r \to +0}I_{++}=0$ and if
$\chi=0$ then
$$
\lim_{r \to +0}I_{++}=\frac{1}{2\pi
i}\int_{\gamma_{e^{\tau/2},\pi/2} ^\pm}\frac{1}{z}\left( \frac
{1-e^{-\tau}z}{1+e^{-\tau} z} \right)^{\Delta t_{1,2}} \left(
\frac{z}{e^{\tau/2}} \right)^{-(x_1+x_2)}dz,
$$
where we pick $\gamma_{e^{\tau/2},\pi/2} ^+$ if $\Delta t_{1,2}
\geq 0$ and $\gamma_{e^{\tau/2},\pi/2} ^-$ otherwise.
\end{claim4}

\begin{claim5}
If $\chi>0$ then $\lim_{r \to +0}I_{--}=0$ and if
$\chi=0$ then
$$
\lim_{r \to +0}I_{--}=\frac{1}{2\pi
i}\int_{\gamma_{e^{\tau/2},\pi/2} ^\pm}\frac{1}{z}\left( \frac
{1-e^{-\tau}z}{1+e^{-\tau} z} \right)^{\Delta t_{3,4}} \left(
\frac{z}{e^{\tau/2}} \right)^{(x_3+x_4)}dz,
$$
where we pick $\gamma_{e^{\tau/2},\pi/2} ^+$ if $\Delta t_{3,4} \geq 0$
and $\gamma_{e^{\tau/2},\pi/2} ^-$ otherwise.
\end{claim5}

We postpone the proof of these claims and proceed with the proof of Theorem \ref{limcorfun}.

a) If $\chi>0$ then the part of the Pfaffian coming
from $++$ and $--$ blocks (two $n \times n$ blocks on the main
diagonal) will not contribute to the limit. This is because every term
in the Pfaffian contains equally many $K_{++}$
and $K_{--}$ factors. Now, by (\ref{constI++I--}) we have that
$K_{++}^{1,2} \cdot K_{--}^{3,4}=\text{const}(-1)^{x_2+x_3} \cdot I_{++}I_{--}$ and then by Claim 4 and Claim 5 we have that $K_{++}^{1,2}\cdot K_{--}^{3,4} \to 0$ since $I_{++} \to 0$ and $I_{--} \to
0$.

This means that the Pfaffian reduces to the determinant of $+-$ block, because
$
\text{Pf}\left[\begin{array}{cc}
0&A\\
-A&0
\end{array}\right]=\det A.
$
By Claim 1 we have that
\begin{eqnarray*}
K_{+-}^{1,2}&=&e^{-\tau(x_1-x_2)}\frac{1}{2\pi
i}\int_{\gamma_{e^{-\tau/2},\theta} ^\pm} \left( \frac {1-z}{1+ z}
\right)^{\Delta t_{1,2}} \frac{1}{z^{\Delta x_{1,2}+1}}dz,
\end{eqnarray*}
and it is easily verified that $e^{\tau(\dots)}$ prefactors
cancel out in the determinant.

Thus, if $\chi>0$ then for
$$
rt_i \to \tau, \;\;\; rx_i \to \chi \;\;\; \text{as } r \to +0,
$$
$$ t_i-t_j= \Delta t_{ij}=\text{const}, \;\;\; x_i-x_j= \Delta x_{ij}=
\text{const},
$$
we have that
$$
\rho(X) \to \det[K(i,j)]_{i,j=1}^n,
$$
where $K(i,j)$ is given in the statement of the theorem.

b) Now, if $\chi=0$ and
$x_i$ does not depend on $r$,
then by Claims 1, 4 and 5 we have that
\begin{eqnarray*}
K_{++}^{1,2}&=&(-1)^{x_2}e^{-\tau(x_1+x_2)/2}\frac{1}{2\pi
i}\int_{\gamma_{e^{\tau/2},\theta} ^\pm} \frac{1}{z} \left( \frac
{1-e^{-\tau}z}{1+e^{-\tau} z} \right)^{\Delta t_{1,2}} \left(
\frac{z}{e^{\tau/2}} \right)^{-(x_1+x_2)}dz\\
&=&(-1)^{x_2}e^{-\tau(x_1+x_2)}\frac{1}{2\pi
i}\int_{\gamma_{e^{-\tau/2},\theta}^\pm} \left( \frac {1-z}{1+ z}
\right)^{\Delta t_{1,2}} \frac{1}{z^{x_1+x_2+1}}dz,
\end{eqnarray*}

\begin{eqnarray*}
K_{--}^{3,4}&=&(-1)^{x_3}e^{\tau(x_3+x_4)/2}\frac{1}{2\pi
i}\int_{\gamma_{e^{\tau/2},\theta} ^\pm} \frac{1}{z} \left( \frac
{1-e^{-\tau}z}{1+e^{-\tau} z} \right)^{\Delta t_{3,4}} \left(
\frac{z}{e^{\tau/2}} \right)^{x_3+x_4}dz\\
&=&(-1)^{x_3}e^{\tau(x_3+x_4)}\frac{1}{2\pi
i}\int_{\gamma_{e^{-\tau/2},\theta} ^\pm} \left( \frac {1-z}{1+ z}
\right)^{\Delta t_{3,4}} \frac{1}{z^{-(x_3+x_4)+1}}dz
\end{eqnarray*}
and
\begin{eqnarray*}
K_{+-}^{1,2}&=&e^{-\tau(x_1-x_2)}\frac{1}{2\pi
i}\int_{\gamma_{e^{-\tau/2},\theta} ^\pm} \left( \frac {1-z}{1+ z}
\right)^{\Delta t_{1,2}} \frac{1}{z^{(x_1-x_2)+1}}dz.
\end{eqnarray*}

As before, it is easily verified that $e^{\tau(\dots)}$ prefactors
cancel out in the Pfaffian.

Hence, if $\chi=0$ then for
$$
rt_i \to \tau, \;\;\; rx_i \to \chi \;\;\; \text{as } r \to +0,
$$
$$ t_i-t_j= \Delta t_{ij}=\text{const}, \;\;\; x_i-x_j= \Delta x_{ij}=
\text{const},\;\;\; x_i=\text{const}
$$
we have that
$$
\rho(X) \to \Pf[M(i,j)]_{i,j=1}^{2n},
$$
where $M(i,j)$ is given in the statement of the theorem.

It remains to prove Claims 1 through 5.

\begin{proof}(Claim 1)
In order to compute the limit of $I_{+-}$ we focus on its
exponentially large term. Since $(z+w)/2zw(z-w)$ remains bounded
away from $z=w$, $z=0$ and $w=0$, the exponentially large term comes
from
\begin{equation}\lb{J1overJ2}
\frac{J(t_1,z)}{J(t_2,w)} \left( \frac{z}{e^{\tau/2}} \right)^{-x_1}
\left(\frac{w}{e^{\tau/2}}\right)^{x_2}=\frac{\exp\left[{ \log
J(t_1,z)-x_1(\log z-\tau/2)}\right]}{\exp{\left[ \log
J(t_2,w)-x_2(\log w-\tau/2)\right]}}.
\end{equation}
To determine a behavior of this term we need to know the asymptotics
of $\log(z;q)_\infty$ when $r\to+0$. Recall that the dilogarithm
function is defined by
$$ \dilog(z)=\sum\limits_{n=1}^\infty
\frac{(1-z)^n}{n^2}, \;\;\;|1-z| \leq1,
$$
with the analytic continuation given by
$$
\dilog (1-z)=\int_1^z \frac{\log(t)}{1-t}dt,
$$
with the negative real axis as a branch cut. Then
$$
\log(z;q)_\infty =- \frac{1}{r}\dilog
(1-z)+O(\operatorname{dist}(1,{rz|0\leq r\leq1})^{-1}),\;\;\;r\to+0.
$$
(see e.g. \cite{B}). Hence, (\ref{J1overJ2}) when $r\to+0$ behaves
like
%
%
$$
\exp{\frac1r \left[S(z,\tau, \chi)-S(w,\tau, \chi)\right]}+O(1),
$$
where \begin{eqnarray*} S(z,\tau, \chi)&=&-\dilog
(1+e^{-\tau}z)-\dilog
(1-\frac{1}{z})+\dilog(1-e^{-\tau}z)\\&&+\dilog(1+\frac{1}{z})-\chi(\log
z-\tau/2).
\end{eqnarray*}
The real part of $S(z,\tau,\chi)$ vanishes for $|z|=e^{\tau/2}$. We
want to find the way it changes when we move from the circle
$|z|=e^{\tau/2}$. For that we need to find
$$
\frac{d}{dR}\text{Re}S(z,\tau,\chi).
$$
On the circle, $\text{Re} S(z,\tau,\chi)=0$ implies that the derivative in the tangent direction vanishes, i.e.
$$
y\frac{d\text{Re}S}{dx}-x\frac{d\text{Re}S}{dy}=0.
$$
The Cauchy-Riemann equations on $|z|=e^{\tau/2}$ then yield
$$
R\frac {d}{dR}\text{Re} S(z, \tau, \chi)=x\frac{d\text{Re}S}{dx}+y\frac{d\text{Re}S}{dy}=z \frac {d}{dz}S(z,\tau, \chi).
$$
Simple calculation gives
$$
z \frac{d}{dz}S(z,\tau,\chi)=-\chi+\log
\frac{(1+e^{-\tau}z)(z+1)}{(1-e^{-\tau}z)(z-1)}
$$
which implies that
$$z \frac{d}{dz}S(z,\tau, \chi)=-\chi+\log \left|
\frac{z+1}{z-1} \right|^2,\;\;\;{\text {for }}|z|=e^{\tau /2}.
$$
Then
\begin{equation}\lb{condi}
z \frac{d}{dz}S(z,\tau, \chi)>0\;\;\;{\rm
iff}\;\;\;e^{\chi/2}<\left| \frac{z+1}{z-1}
\right|,\;\;\;\text{for }|z|=e^{\tau/2}.
\end{equation}
One easily computes that $e^{\chi/2}=\left| \frac{z+1}{z-1}
\right|$ is a circle with center $\frac{R^2+1}{R^2-1}$ and radius
$\frac{2R}{R^2-1}$ where $R=e^{\chi/2}.$ If
\begin{equation}\lb{uslovtheta}
\frac{(e^{\tau}+1)(e^{\chi}-1)}{2e^{\tau/2}{(e^{\chi}+1)}}>1
\end{equation}
then this circle does not intersect $|z|=e^{\tau/2}$ and
$\frac{d}{dR}\text{Re}S(z,\tau,\chi)$ is negative for every point
$z$ such that $|z|=e^{\tau/2}$. Otherwise, this circle intersects
$|z|=e^{\tau/2}$ at $z=e^{\tau/2}e^{\pm i\theta}$ where $\theta$ is
given with (\ref{theta}). Thus, the sign of
$\frac{d}{dR}\text{Re}S(z,\tau,\chi)$ changes at these points.

Let $\gamma _z$ and $\gamma_w$ be as in Figure \ref{Contours+-}. We
pick them in such a way that the real parts of $S(z,\tau,\chi)$
and $S(w,\tau,\chi)$  are negative everywhere on these contours
except at the two points with the argument equal to
$\pm\theta$ (if (\ref{uslovtheta}) is satisfied they are negative at these points too). This is possible by (\ref{condi}).
\begin{figure}[htp!]
\includegraphics[height=7cm]{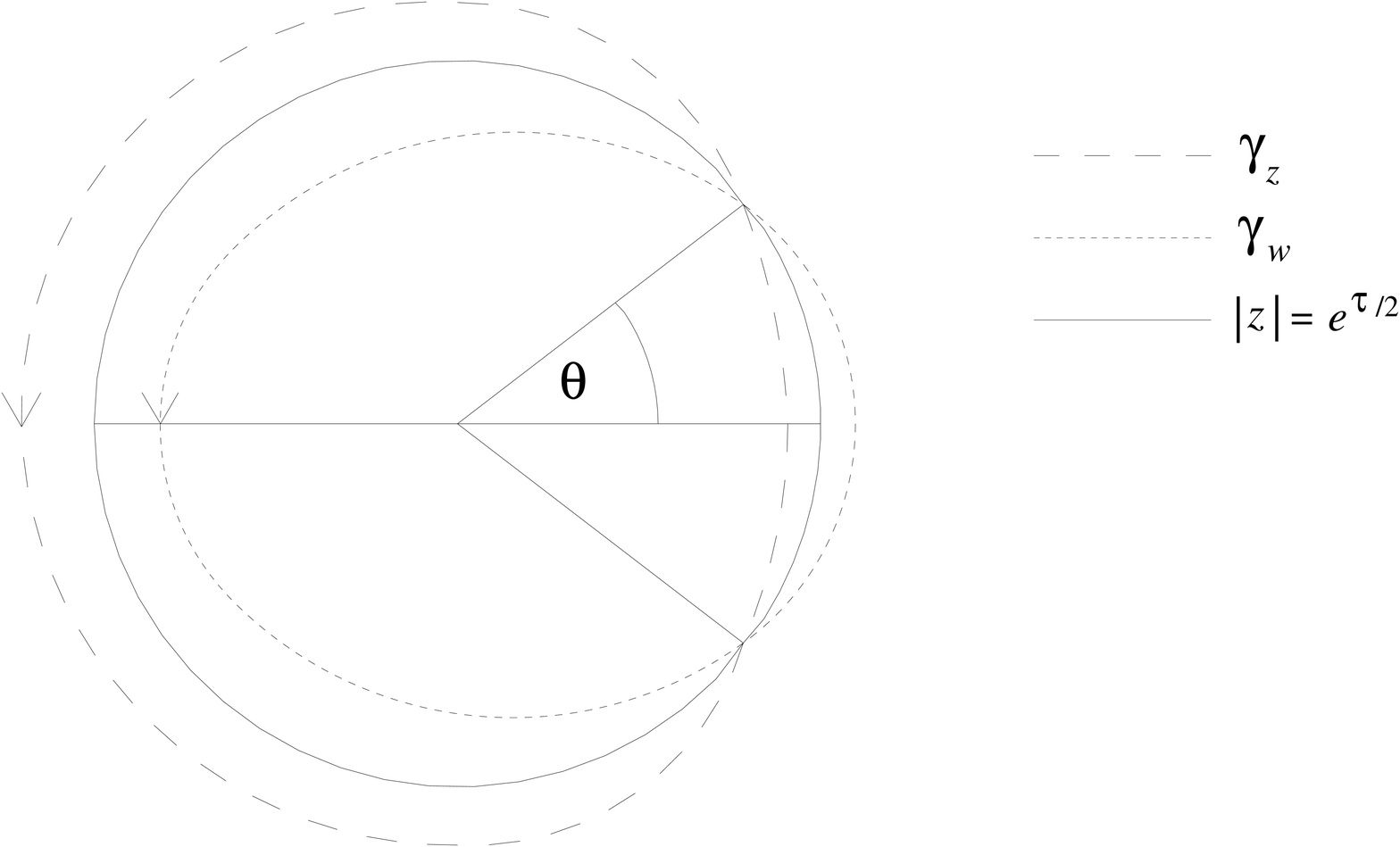}
\caption{Contours $\gamma _z$ and $\gamma_w$} \label{Contours+-}
\end{figure}

The reason we are introducing $\gamma_z$ and $\gamma_w$ is that we want to deform the contours
in $I_{+-}$ to these two and then use that
\begin{equation*}
\int_\gamma e^{\frac{S(z)}{r}}dz \to 0, \;\;\; r\to +0.
\end{equation*}
whenever $S$ has a negative real part for all but finitely many
points on $\gamma$ . When deforming contours we will need to include
the contribution coming from the poles and that way the integral
$I_{+-}$ will be a sum of integrals where the first one vanishes as
$r \to +0$ and the other nonvanishing one comes as a residue.

Let $f(z,w)$ be the integrand in $I_{+-}$. We consider two cases a) $t_1 \geq t_2$ and b)
$t_1<t_2$ separately.

Case a) $t_1 \geq t_2$. We omit the integrand $f(z,w)dzdw$ in the
formulas below.

\begin{eqnarray*}
I_{+-}&=&\frac{1}{(2\pi i)^2} \int \limits
_{|z|=(1+\epsilon)e^{\tau/2}}\int
\limits_{|w|=(1-\epsilon)e^{\tau/2}}=\frac{1}{(2\pi i)^2} \int  \limits_{\gamma _z}\int\limits
_{|w|=(1-\epsilon)e^{\tau/2}}\\
&=&\frac{1}{(2\pi i)^2} \left[ -\int  _{\gamma^- _{z,\theta}}\int
_{|w|=(1-\epsilon)e^{\tau/2}} + \int_{\gamma^+
_{z,\theta}}\int_{|w|=(1-\epsilon)e^{\tau/2}}  \right]\\
&=&\frac{1}{(2\pi i)^2} \left[- \int_{\gamma^-
_{z,\theta}}\int_{\gamma_w}
 + \int_{\gamma^+
_{z,\theta}}\int_{\gamma_w} -2 \pi i \int_{\gamma^+_{z,\theta}}   \text{Res}_{w=z} f(z,w)dz\right]\\
&=&\frac{1}{(2\pi i)^2} \left[ \int_{\gamma _z}\int_{\gamma_w}
 -2 \pi i\int_{\gamma^+
_{z,\theta}}  \text{Res}_{w=z} f(z,w)dz\right]\\
&=&I_1-\frac{1}{2\pi i}\int_{\gamma_{e^{\tau/2},\theta}^
+}\frac{2z}{-2z^2}\frac{J(t_1,z)}{J(t_2,z)} \cdot \left(
\frac{z}{e^{\tau/2}}\right) ^{-x_1}\cdot
\left(\frac{z}{e^{\tau/2}}\right)^{x_2}dz\\
&=&I_1+\frac{1}{2\pi i}\int_{\gamma_{e^{\tau/2},\theta}^
+}\frac{1}{z}\frac{J(t_1,z)}{J(t_2,z)} \cdot \left(
\frac{z}{e^{\tau/2}}\right) ^{-\Delta x_{1,2}}dz.
\end{eqnarray*}

The first integral denoted with $I_1$ vanishes as $r \to +0$, while
the other one goes to the integral from our claim since
$$
\lim_{r \to
+0}\frac{J(t_1,z)}{J(t_2,z)}=\left(\frac{1-e^{-\tau}z}{1+e^{-\tau}z}\right)^{\Delta
t_{1,2}}.
$$

Case b) $t_1<t_2$ is handled similarly
\begin{eqnarray*}
 I_{+-}&=& \frac{1}{(2\pi i)^2}
\int \limits_{|z|=(1-\epsilon)e^{\tau/2}}\int \limits_{|w|=(1+\epsilon)e^{\tau/2}}=\frac{1}{(2\pi i)^2} \int\limits_{\gamma
_z}\int\limits_{|w|=(1+\epsilon)e^{\tau/2}}\\
&=&\frac{1}{(2\pi i)^2} \left[ \int_{\gamma _z}\int_{\gamma_w}
 - 2 \pi i\int_{\gamma_{e^{\tau/2},\theta}^-
} \text{Res} _{w=z}f(z,w)dz\right]\\
&=&I_2-\frac{1}{2\pi i}\int_{\gamma_{e^{\tau/2},\theta}^
-}\frac{2z}{-2z^2}\frac{J(t_1,z)}{J(t_2,z)} \cdot \left(
\frac{z}{e^{\tau/2}}\right) ^{-x_1}\cdot
\left(\frac{z}{e^{\tau/2}}\right)^{x_2}dz\\
&=&I_2+\frac{1}{2\pi i}\int_{\gamma_{e^{\tau/2},\theta}
^-}\frac{1}{z}\frac{J(t_1,z)}{J(t_2,z)} \cdot \left(
\frac{z}{e^{\tau/2}}\right) ^{-\Delta x_{1,2}}dz.
\end{eqnarray*}
\end{proof}

%
%
\begin{proof}(Claim 2)
The exponentially large term of $I_{++}$ comes from
$$
\frac{J(t_1,z)}{J(t_2,w)} \left( \frac{z}{e^{\tau/2}} \right)^{-x_1}
\left(\frac{w}{e^{\tau/2}}\right)^{-x_2}=\frac{\exp\left[{\log
J(t_1,z)-x_1(\log z-\tau /2)}\right]}{\exp\left[\log
J(t_2,w)+x_2(\log w-\tau/2)\right]}
$$
that when $r \to +0$ behaves like
$$
\exp{\frac1r \left[S(z,\tau, \chi)+T(w,\tau, \chi)\right]}+O(1),
$$
where
\begin{eqnarray*}
S(z,\tau, \chi)&=&-\dilog (1+e^{-\tau}z)-\dilog
(1-\frac{1}{z})+\dilog(1-e^{-\tau}z)\\&&+\dilog(1+\frac{1}{z})-\chi
(\log z-\tau/2)
\end{eqnarray*}
and
\begin{eqnarray*}
T(w,\tau,\chi)&=&\dilog(1+e^{-\tau
}w)+\dilog(1-\frac{1}{w})-\dilog(1-e^{-\tau
}w)\\&&-\dilog(1+\frac{1}{w})-\chi(\log w-\tau/2).
\end{eqnarray*}

Real parts of $S(z,\tau, \chi)$ and $T(w,\tau, \chi)$ vanish for
$|z|=e^{\tau/2}$ and $|w|=e^{\tau/2}$, respectively. To see the way
they change when we move from these circles we need $\frac
{d}{dR}\text{Re}S(z,\tau,\chi)$
 and $\frac
{d}{dR}\text{Re}T(w,\tau,\chi)$. For $|z|=e^{\tau/2}$ and
$|w|=e^{\tau/2}$  they are equal to $\frac{z}{R} \frac
{d}{dz}S(z,\tau, \chi)$
 and $ \frac{w}{R}\frac {d}{dz}T(w,\tau, \chi)$, respectively.


The needed estimate for $S$ is (\ref{condi}) above.

Similarly,
$$
w \frac{d}{dw}T(w,\tau, \chi)>0\;\;\;{\rm
iff}\;\;\;e^{\chi/2}<\left| \frac{w-1}{w+1} \right|,\;\;\;{\text{for }}|w|=e^{\tau/2}.
$$

Hence, if (\ref{uslovtheta}) holds
then both $\frac{d}{dR}\text{Re}S(z,\tau,\chi)$ and
$\frac{d}{dR}\text{Re}T(z,\tau,\chi)$ are negative for every point
$z$ such that $|z|=e^{\tau/2}$. Otherwise, the sign of
$\frac{d}{dR}\text{Re}S(z,\tau,\chi)$ changes at $z=e^{\tau/2}e^{\pm
i\theta}$, while the sign of $\frac{d}{dR}\text{Re}T(w,\tau,\chi)$
changes at $w=e^{\tau/2}e^{\pm i(\pi-\theta)}$.

We deform the contours in $I_{++}$ to $\gamma_z$ and $\gamma_w$
shown in Figure \ref{Contours++} because the real parts of $S$ and
$T$ are negative on these contours (except for finitely many
points).
\begin{figure}[htp!]
\includegraphics[height=7cm]{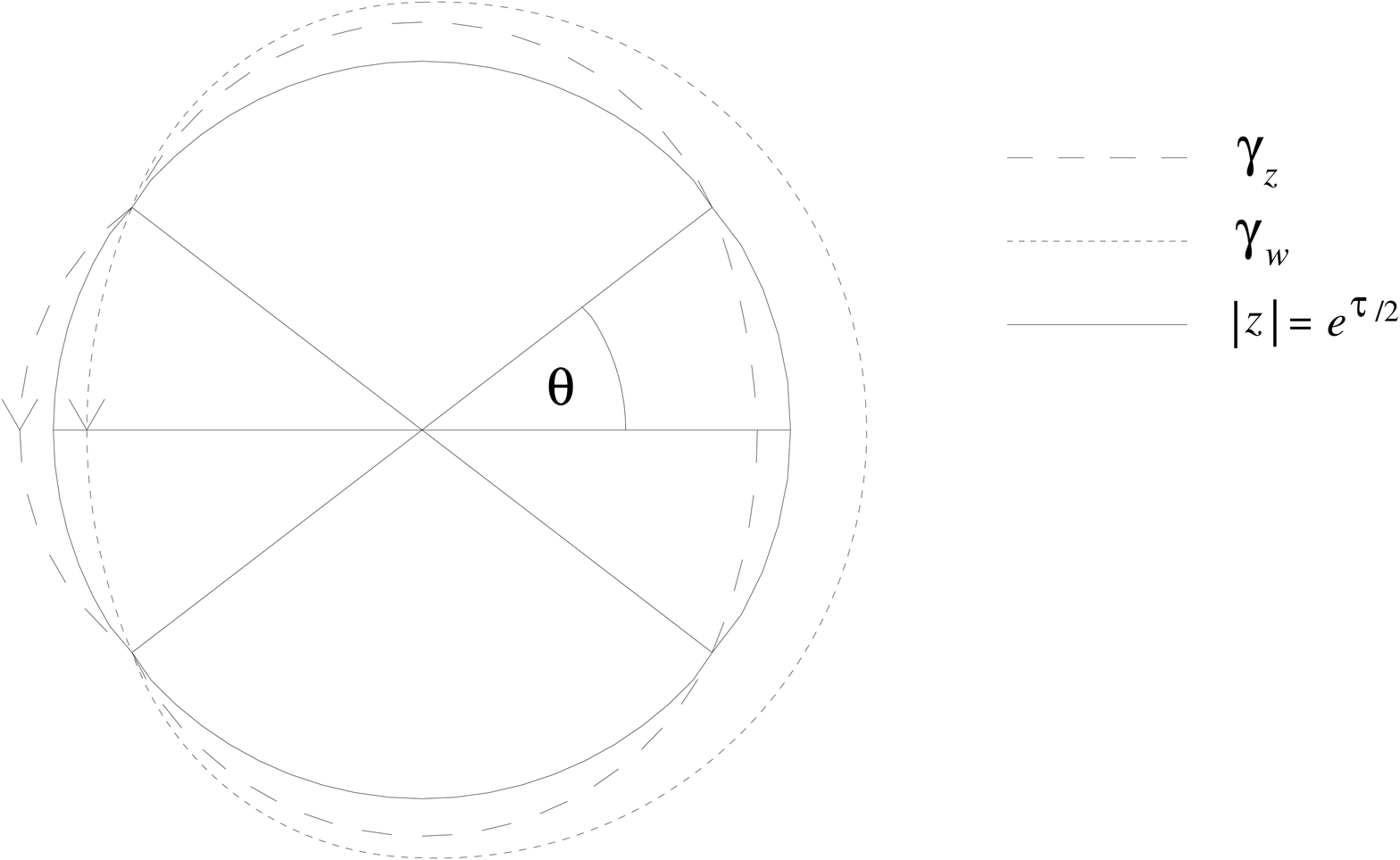}
\caption{Contours $\gamma_z$ and $\gamma_w$} \label{Contours++}
\end{figure}

As before, we distinguish two cases a) $t_1 \geq t_2$ and b)
$t_1<t_2$.

For a) $t_1 \geq t_2$ (where we again omit $f(z,w)dzdw$)
\begin{eqnarray*}
I_{++}&=&\frac{1}{(2\pi i)^2} \int
\limits_{|z|=(1+\epsilon)e^{\tau/2}}\int
\limits_{|w|=(1-\epsilon)e^{\tau/2}}=\frac{1}{(2\pi i)^2} \int\limits_{\gamma
_z}\int\limits_{|w|=(1-\epsilon)e^{\tau/2}}\\
&=&\frac{1}{(2\pi i)^2} \left[- \int_{\gamma^{-}
_{z,\pi-\theta}}\int _{|w|=(1-\epsilon)e^{\tau/2}} +
\int_{\gamma^{+}
_{z,\pi-\theta}}\int_{|w|=(1-\epsilon)e^{\tau/2}}  \right]\\
&=&\frac{1}{(2\pi i)^2} \left[ -\int_{\gamma^{-}
_{z,\pi-\theta}}\int _{\gamma_w} + \int_{\gamma^{+}
_{z,\pi-\theta}}\int_{\gamma_w}  - 2 \pi i \int_{\gamma^{+}
_{z,\pi-\theta}}  \text{Res}_{w=z} f(z,w)dz\right]\\
&=&\frac{1}{(2\pi i)^2} \left[ \int_{\gamma _z}\int_{\gamma_w} - 2
\pi i \int_{\gamma_{e^{\tau/2},\pi-\theta}^+
} \text{Res}_{w=z} f(z,w)dz\right]\\
&=&I_1-\frac{1}{2\pi i}\int_{\gamma_{e^{\tau/2},\pi-\theta}^+
}\frac{2z}{-2z^2}\frac{J(t_1,z)}{J(t_2,z)} \cdot \left(
\frac{z}{e^{\tau/2}}\right) ^{-x_1}\cdot
\left(\frac{z}{e^{\tau/2}}\right)^{-x_2}dz.\\
\end{eqnarray*}
For the same reason as in the proof of Claim 1 we have that $I_1 \to
0$ when $r \to +0$. Thus, $\lim_{r \to +0} I_{++}=\lim_{r \to +0}
I_2$, where
\begin{eqnarray*}
I_2&=&\frac{1}{2\pi i}\int_{\gamma_{e^{\tau/2},\pi-\theta}^+
}\frac{1}{z}\frac{J(t_1,z)}{J(t_2,z)} \cdot \left(
\frac{z}{e^{\tau/2}}\right) ^{-x_1}\cdot
\left(\frac{z}{e^{\tau/2}}\right)^{-x_2}dz\\
&=&\frac{1}{2\pi i}\int_{\gamma_{e^{\tau/2},\theta}^-
}\frac{1}{-z}\frac{J(t_1,-z)}{J(t_2,-z)} \cdot \left(
\frac{-z}{e^{\tau/2}}\right) ^{-x_1}\cdot
\left(\frac{-z}{e^{\tau/2}}\right)^{-x_2}dz\\
&=&(-1)^{x_1+x_2+1}\cdot \frac{1}{2 \pi i} \int_{
\gamma_{e^{\tau/2},\theta}^-}
\frac{1}{z}\frac{J(t_2,z)}{J(t_1,z)}\left( \frac{z}{e^ {\tau
/2}}\right)^{-(x_1+x_2)}dz.
\end{eqnarray*}

For b) $t_1<t_2$
\begin{eqnarray*}
 I_{++}&=& \frac{1}{(2\pi i)^2}
\int \limits_{|z|=(1-\epsilon)e^{\tau/2}}\int \limits_{|w|=(1+\epsilon)e^{\tau/2}}=\frac{1}{(2\pi i)^2} \int\limits_{\gamma
_z}\int\limits_{|w|=(1+\epsilon)e^{\tau/2}}\\
&=&\frac{1}{(2\pi i)^2} \left[ \int_{\gamma _z}\int_{\gamma_w}
-2 \pi i\int_{\gamma_{e^{\tau/2},\pi-\theta}^-
}  \text{Res}_{w=z} f(z,w)dz\right]\\
&=&I_3-\frac{1}{2\pi i}\int_{\gamma_{e^{\tau/2},\pi-\theta}^-
}\frac{2z}{-2z^2}\frac{J(t_1,z)}{J(t_2,z)} \cdot \left(
\frac{z}{e^{\tau/2}}\right) ^{-x_1}\cdot
\left(\frac{z}{e^{\tau/2}}\right)^{-x_2}dz.
\end{eqnarray*}

Thus, $\lim_{r \to +0} I_{++}=\lim_{r \to +0} I_4$, where
\begin{eqnarray*}
I_4&=&\frac{1}{2\pi i}\int_{\gamma_{e^{\tau/2},\pi-\theta}^-
}\frac{1}{z}\frac{J(t_1,z)}{J(t_2,z)} \cdot \left(
\frac{z}{e^{\tau/2}}\right) ^{-x_1}\cdot
\left(\frac{z}{e^{\tau/2}}\right)^{-x_2}dz\\
&=&\frac{1}{2\pi i}\int_{\gamma_{e^{\tau/2},\theta}^+
}\frac{1}{-z}\frac{J(t_1,-z)}{J(t_2,-z)} \cdot \left(
\frac{-z}{e^{\tau/2}}\right) ^{-x_1}\cdot
\left(\frac{-z}{e^{\tau/2}}\right)^{-x_2}dz\\
&=&(-1)^{x_1+x_2+1}\cdot \frac{1}{2 \pi i} \int_{
\gamma_{e^{\tau/2},\theta} ^+}
\frac{1}{z}\frac{J(t_2,z)}{J(t_1,z)}\left( \frac{z}{e^ {\tau
/2}}\right)^{-(x_1+x_2)}dz.
\end{eqnarray*}

Therefore, we conclude that
$$
\lim _{r \to +0} I_{++}= \lim _{r \to +0} I,
$$
with
$$
I=(-1)^{x_1+x_2+1}\cdot \frac{1}{2 \pi i} \int_{
\gamma_{e^{\tau/2},\theta} ^\mp}
\frac{1}{z}\frac{J(t_2,z)}{J(t_1,z)}\left( \frac{z}{e^ {\tau
/2}}\right)^{-(x_1+x_2)}dz,
$$
where we choose $ \gamma_{e^{\tau/2},\theta} ^-$ if $t_1 \geq t_2$
and $\gamma_{e^{\tau/2},\theta} ^+$ otherwise.
\end{proof}

%
\begin{proof}(Claim 3)
The exponentially large term of $I_{--}$ comes from
$$
\frac{J(t_3,z)}{J(t_4,w)} \left( \frac{z}{e^{\tau/2}} \right)^{x_3}
\left(\frac{w}{e^{\tau/2}}\right)^{x_4}=\frac{\exp\left[{\log
J(t_3,z) +x_3(\log z-\tau /2)}\right]}{\exp\left[\log
J(t_4,w)-x_4(\log w-\tau/2)\right]}
$$
that when $r \to +0$ behaves like
$$
\exp{\frac1r \left[S(z,\tau, \chi)+T(w,\tau, \chi)\right]}+O(1),
$$
where
\begin{eqnarray*}
S(z,\tau, \chi)&=&-\dilog (1+e^{-\tau}z)-\dilog
(1-\frac{1}{z})+\dilog(1-e^{-\tau}z)\\&&+\dilog(1+\frac{1}{z})+\chi
(\log z-\tau/2)
\end{eqnarray*}
and
\begin{eqnarray*}
T(w,\tau,\chi)&=&\dilog(1+e^{-\tau
}w)+\dilog(1-\frac{1}{w})-\dilog(1-e^{-\tau
}w)\\&&-\dilog(1+\frac{1}{w})+\chi(\log w-\tau/2).
\end{eqnarray*}

As before, real parts of $S(z,\tau, \chi)$ and $T(w,\tau, \chi)$ vanish for
$|z|=e^{\tau/2}$ and $|w|=e^{\tau/2}$, respectively.

Since
$$
z \frac{d}{dz}S(z,\tau, \chi)>0\;\;\;{\rm
iff}\;\;\;e^{\chi/2}>\left| \frac{z-1}{z+1} \right|\;\;\;\text{for }|z|=e^{\tau/2}
$$
and
$$
w \frac{d}{dw}T(w,\tau, \chi)>0\;\;\;{\rm
iff}\;\;\;e^{\chi/2}>\left| \frac{w+1}{w-1} \right|\;\;\;\text{for }|w|=e^{\tau/2},
$$
we deform contours in $I_{--}$ to $\gamma_z$ and $\gamma_w$ shown in
Figure \ref{Contours--}.
\begin{figure}[htp!]
\includegraphics[height=7cm]{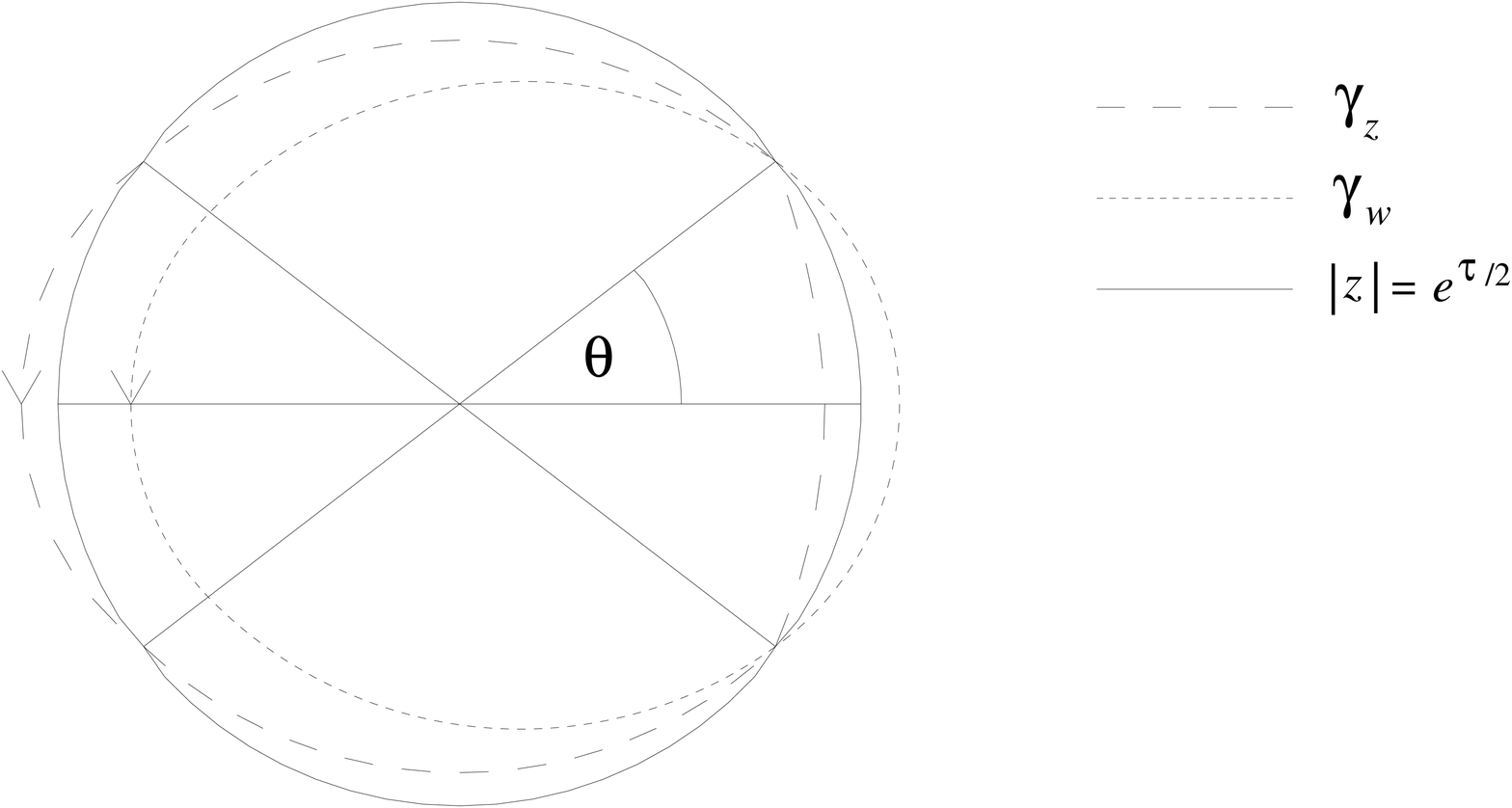}
\caption{Contours $\gamma_z$ and $\gamma_w$} \label{Contours--}
\end{figure}

Using the same reasoning as in the proof of Claim 1 and 2 we get that
$$
\lim _{r \to +0} I_{--}= \lim _{r \to +0} I,
$$
with
$$
I=\frac{1}{2 \pi i} \int_{\gamma_{e^{\tau/2},\theta} ^\pm}
\frac{1}{z}\frac{J(t_3,z)}{J(t_4,z)}\left( \frac{z}{e^ {\tau
/2}}\right)^{x_3+x_4}dz,
$$
where we choose $\gamma_{e^{\tau/2},\theta} ^+$ if $t_3\geq t_4$ and $\gamma_{e^{\tau/2},\theta} ^-$ otherwise.
\end{proof}


\begin{proof}(Claim 4)
We start from the integral on the right hand side in Claim 2.

Because
\begin{equation}\lb{JJ}
\lim_{r \to +0}\frac {J(z,t_1)}{J(z,t_2)}=\left( \frac{1-e^{-\tau
}z}{1+e^{-\tau}z} \right)^{\Delta t_{1,2}},
\end{equation}
we focus on
$$
\lim_{r \to +0}\left(\frac{z}{e^{\tau/2}}\right)^{-(x_1+x_2)}=\exp\left[\frac{S(z,\tau,\chi)}{r}\right],
$$
where
$$
S(z,\tau, \chi)=-2\chi \log z+\chi \tau.
$$

Assume $\chi >0$.
Then $\text{Re}S(z,\tau,\chi)>0$ if and only if
$|z|<e^{\tau/2}$.

For $\Delta t_{1,2} \geq 0$ we deform  the contour $\gamma
^-_{e^{\tau/2},\theta}$ to $\gamma _z$ (see Figure \ref{Con++}).
Using the same argument as in the proof of Claim 1 we get that
$\lim_{r \to +0} I_{++}=0$ for $\Delta t_{1,2} \geq 0$. A similar argument
can be given for $\Delta t_{1,2}<0$.
\begin{figure}[htp!]
\includegraphics[height=7cm]{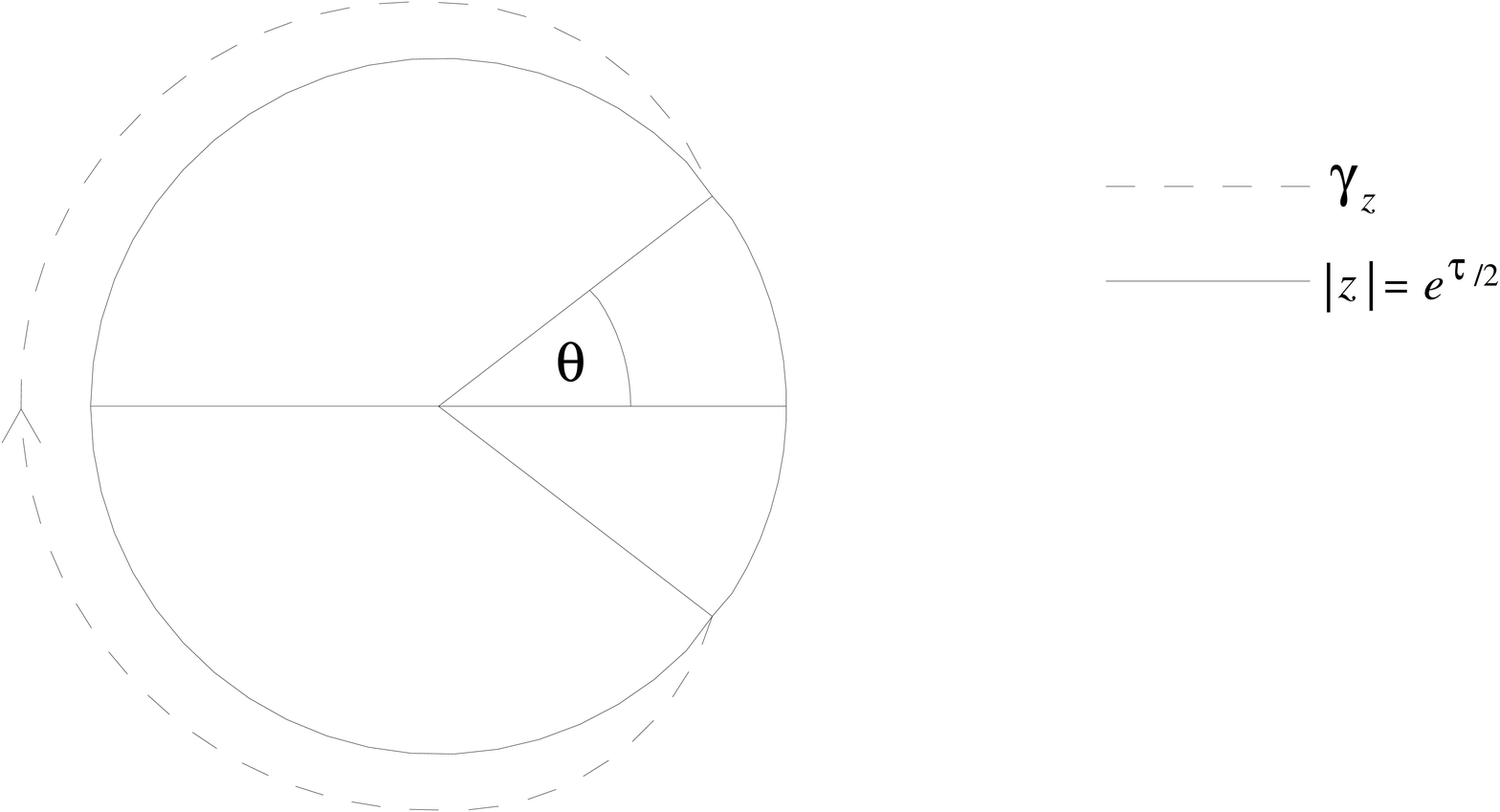}
\caption{Contour $\gamma_z$} \label{Con++}
\end{figure}

For $\chi=0$ the claim follows directly from Claim 2 changing $z \mapsto -z$ and using (\ref{JJ}).
\end{proof}

\begin{proof}(Claim 5)
The proof is similar to the proof of Claim 4.
%
\end{proof}

The proof of Theorem \ref{limcorfun} is now complete.
\end{proof}

\bigskip

\section{Appendix} \lb{appendix}

\subsection{Strict partitions} \lb{}

A strict partition is a sequence of strictly decreasing integers
such that only finitely many of them are nonzero. The nonzero
integers are called parts. Let throughout this section
$\lambda=(\lambda_1,\lambda_2, \dots)$ and $\mu=(\mu_1,\mu_2,\dots)$
be strict partitions.

The empty partition is $\emptyset=(0,0,\dots)$.

The weight of $\lambda$ is $|\lambda|=\sum\lambda_i$.

The length of $\lambda$ is $l(\lambda)= \# \text{ of parts of }
\lambda$.

The diagram of $\lambda$ is the set of points $(i,j) \in \bbZ^2$
such that $1 \leq j \leq \lambda_i$. The points of the diagram are
represented by $1 \times 1$ boxes. The shifted diagram of $\lambda$
is the set of points $(i,j) \in \bbZ^2$ such that $i \leq j \leq
\lambda_i+i-1$.

\begin{figure}[htp!]
\includegraphics[height=2cm]{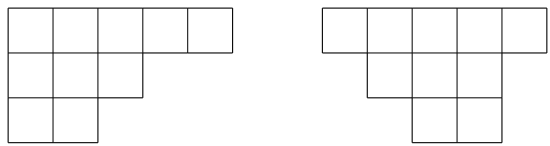}
\caption{Diagram and shifted diagram of $(5,3,2)$}
\label{ApendixPPFin}
\end{figure}

The diagram and the shifted diagram of $\lambda=(5,3,2)$ are shown
in Figure \ref {ApendixPPFin}.

A partition $\mu$ is a subset of $\lambda$ if $\mu_i \leq \lambda_i$
for every $i$. We write $\mu \subset \lambda$ or $\lambda \supset
\mu$ in that case. This means that the diagram of $\lambda$ contains
the diagram of $\mu$.

If $\lambda \supset \mu$, the (shifted) skew diagram $\lambda-\mu$
is the set theoretic difference of the (shifted) diagrams of
$\lambda$ and $\mu$.

\begin{figure}[htp!]
\includegraphics[height=2cm]{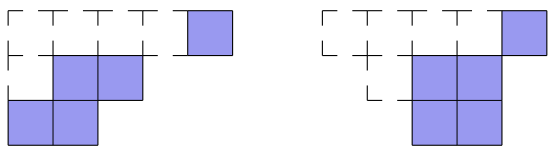}
\caption{Skew diagram and shifted skew diagram of $(5,3,2)-(4,1)$}
\label{ApendixPPSkewFin}
\end{figure}

If $\lambda=(5,3,2)$ and $\mu=(4,1)$ then $\lambda \supset \mu$ with
the skew and the shifted skew diagram $\lambda-\mu$ as in Figure
\ref{ApendixPPSkewFin}.

A skew diagram is a horizontal strip if it does not contain more
than one box in each column. The given example is not a horizontal
strip.

A connected part of a (shifted) skew diagram that contains no $2
\times 2$ block of boxes is called a border strip. The skew diagram
of the example above has two border strips.

If $\lambda \supset \mu$ and $\lambda-\mu$ is a horizontal strip
then we define $a(\lambda - \mu)$ as the number of integers $i \geq
1$ such that the skew diagram $\lambda - \mu$ has a box in the $i$th
column but not in the $(i+1)$st column or, equivalently, as the
number of mutually disconnected border strips of the shifted skew
diagram $\lambda-\mu$.


Let $P$ be a totally ordered set
$$
P=\{1<1'<2<2'< \cdots\}.
$$
We distinguish elements in $P$ as unmarked and marked, the latter
being one with a prime. We use $|p|$ for the unmarked number
corresponding to $p \in P$.

A marked shifted (skew) tableau is a shifted (skew) diagram filled
with row and column nonincreasing elements from $P$ such that any
given unmarked element occurs at most once in each column whereas
any marked element occurs at most once in each row. Examples of a
marked shifted tableau and a marked shifted skew tableau are
$$
\begin{array}{ccccc}
5&3'&2'&1'&1\\
&3&2'&1'&\\
&&1&1
\end{array}\;\;\;\;\;\;\;
\begin{array}{ccccc}
&&&1'&1\\
&&2'&1'&\\
&&1&1
\end{array}
$$

For $x=(x_1,x_2,\dots,x_n)$ and a marked shifted (skew) tableau $T$
we use $x^T$ to denote $x_1^{a_1}x_2^{a_2} \dots x_n^{a_n}$, where
$a_i$ is equal to the number of elements $p$ in $T$ such that
$|p|=i$.

The skew Schur function $Q_{\lambda/\mu}$ is a symmetric function
defined as
\begin{equation}\lb{Qtableau}
Q_{\lambda / \mu} (x_1,x_2, \dots,x_n)=
\begin{cases}
\sum_{T}x^{|T|},&\lambda\supset\mu,\\
0,&\text{otherwise},
\end{cases}
\end{equation}
where the sum is taken over all marked skew shifted tableaux of
shape $\lambda-\mu$ filled with elements $p\in P$ such that $|p|\leq
n$. Skew Schur $P_{\lambda/\mu}$ function is defined as $P_{\lambda/
\mu}=2^{l(\mu)-l(\lambda)}Q_{\lambda/ \mu}$. Also, one denotes
$P_\lambda=P_{\lambda/\emptyset}$ and
$Q_\lambda=Q_{\lambda/\emptyset}$.

This is just one of many ways of defining Schur $P$ and $Q$
functions (see Chapter 3 of \cite{Mac}).

We set
\begin{equation}\lb{Hfrac}
H(x,y)=\prod_{i,j}\frac{1+x_iy_j}{1-x_iy_j}
\end{equation}
and
\begin{equation}\lb{Q}
F(x;z)=\prod_{i=1}^\infty\frac{1+x_iz}{1-x_iz}.
\end{equation}

$F(x;z)$ is denoted with $Q_x(z)$ in \cite{Mac}.

Let $\Lambda$ be the algebra of symmetric functions. A
specialization of $\Lambda$ is an algebra homomorphism $\Lambda \to
\bbC$. If $\rho$ is a specialization of $\Lambda$ then we write
$P_{\lambda/\mu}(\rho)$ and $Q_{\lambda/\mu}(\rho)$ for the images
of $P_{\lambda/\mu}$ and $Q_{\lambda/\mu}$, respectively. Every map
$\rho_x:(x_1,x_2,\dots)\to (a_1,a_2,\dots)$ where $a_i\in \bbC$ and
only finitely many $a_i$'s are nonzero defines a specialization. For
that case the definition (\ref{Qtableau}) is convenient for
determining $Q_{\lambda/\mu}(\rho_x)$. We use $H(\rho_x,\rho_y)$ and
$F(\rho_x;z)$ for the images of (\ref{Hfrac}) and (\ref{Q}) under
$\rho_x \otimes \rho_y$ and $\rho_x$, respectively.

We recall some facts that can be found in Chapter 3 of \cite{Mac}:

\begin{equation} \lb{QP}
H(x,y)=\sum_{\lambda \; {\rm
strict}}Q_\lambda(x)P_\lambda(y)=\prod_{i,j}\frac{1+x_iy_j}{1-x_iy_j},
\end{equation}

\begin{equation} \lb{5.5}
Q_\lambda(x,z)=\sum_{\mu \; {\rm strict}}Q_{\lambda /
\mu}(x)Q_\mu(z),
\end{equation}

\begin{equation} \lb{5.5'}
P_\lambda(x,z)=\sum_{\mu \; {\rm strict}}P_{\lambda /
\mu}(x)P_\mu(z),
\end{equation}

\begin{equation} \lb{5.1}
Q_{\lambda/ \mu}(x)=\sum_{\nu \; {\rm strict}}f_{\mu \nu}^\lambda
Q_\nu(x),
\end{equation}

\begin{equation} \lb{5.2}
P_\mu(x)P_\nu(x)=\sum_{\lambda \; {\rm strict}}f_{\nu \mu}^\lambda
P_\lambda(x),
\end{equation}
for $f^\lambda_{\mu \nu} \in \bbZ$.

\begin{proposition}\lb{QPQP}
\begin{equation*}
\sum_{\lambda \; {\rm strict}}Q_{\lambda / \mu}(x)P_{\lambda /
\nu}(y)=H(x,y)\sum_{\tau \; {\rm strict}}Q_{\nu / \tau}(x)P_{\mu /
\tau}(y)
\end{equation*}
\end{proposition}

\begin{proof}
Let $H=H(x,y)H(x,u)H(z,y)H(z,u)$. Then by (\ref{QP}) we have
\begin{equation*}
H=\sum_{\lambda}Q_{\lambda}(x,z)P_{\lambda}(y,u).
\end{equation*}
We can compute $H$ in two different ways using (\ref{5.5}),
(\ref{5.5'}), (\ref{5.1}) and (\ref{5.2}). One way we get
\begin{eqnarray}
H &=& \sum_{\lambda}Q_{\lambda}(x,z)P_{\lambda}(y,u) =
\sum_{\lambda, \mu, \nu}Q_{\lambda / \mu}(x)Q_{\mu}(z)P_{\lambda /
\nu}(y)P_{\nu}(u)
\nonumber \\
  &=& \sum_{\mu, \nu}\left[\sum_{\lambda}Q_{\lambda / \mu}(x)P_{\lambda / \nu}(y)\right]Q_{\mu}(z)P_{\nu}(u)
  \lb{H1}.
\end{eqnarray}
The other way we get
\begin{eqnarray}
H &=& H(x,y) \sum_{\sigma, \rho,
\tau}Q_{\sigma}(x)P_{\sigma}(u)Q_{\rho}(z)P_{\rho}(y)Q_{\tau}(z)P_{\tau}(u)
\nonumber \\
  &=& H(x,y) \sum_{\sigma, \rho, \tau}Q_{\sigma}(x)P_{\rho}(y)(Q_{\rho}(z)Q_{\tau}(z))(P_{\sigma}(u)P_{\tau}(u)) \nonumber \\
  &=& H(x,y) \sum_{\sigma, \rho, \tau}Q_{\sigma}(x)P_{\rho}(y)\sum _{\mu}2^{-l(\mu)}2^{l(\rho)}2^{l(\tau)}f_{\rho
  \tau}^{\mu}Q_{\mu}(z)\sum_{\nu}f_{\sigma \tau}^{\nu}P_{\nu}(u) \nonumber \\
  &=&H(x,y) \sum_{\tau, \mu, \nu}Q_{\mu}(z)P_{\nu}(u)(\sum _{\rho}2^{l(\tau)}2^{l(\rho)}2^{-l(\mu)}f_{\rho
  \tau}^{\mu}P_{\rho}(y))(\sum_{\sigma}f_{\sigma \tau}^{\nu}Q_{\sigma}(x)) \nonumber \\
 &=&H(x,y) \sum_{\tau, \mu, \nu}Q_{\mu}(z)P_{\nu}(u)P_{\mu / \tau}(y)Q_{\nu / \tau}(x) \nonumber \\
 &=& \sum_{\mu, \nu}\left[H(x,y) \sum_{\tau}P_{\mu / \tau}(y)Q_{\nu / \tau}(x)\right]Q_{\mu}(z)P_{\nu}(u)
 \lb{H2}.
\end{eqnarray}

Now, the proposition follows from (\ref{H1}) and (\ref{H2}).
\end{proof}

We give another proof of this proposition at the end of Section
\ref{Fock}.

\subsection{Fock space associated with strict partitions} \lb{Fock}
We introduce a Fock space associated with strict partitions. We
follow \cite{Mat}. See also \cite{DJKM}.

Let V be a vector space generated by vectors
$$
v_\lambda=e_{\lambda_1} \wedge e_{\lambda_2} \wedge \dots \wedge
e_{\lambda_l},
$$
where $\lambda=(\lambda_1,\lambda_2, \dots, \lambda_l)$ is a strict
partition. In particular, $v_{\emptyset}=1$ and is called the
vacuum.

For every $k \in \bbN_0= \bbN\cup \{0\}$ we define  two (creation
and annihilation) operators $\psi_k$ and $\psi_k^*$. These two
operators are adjoint to each other for the inner product defined by
$$
\langle v_\lambda, v_\mu \rangle= 2^{-l(\lambda)}\delta_{\lambda,
\mu}.
$$

For $k=0$, let
$$
\psi_0 v_\lambda=\psi^{*}_{0}
v_{\lambda}=\frac{(-1)^{l(\lambda)}}{2}v_{\lambda}.
$$

For $k \geq 1$, the operator $\psi_k$ adds $e_k$, on the left while
$\psi^*_k$ removes $e_k$, on the left and divides by 2. More
precisely,
\begin{eqnarray*}
\psi_kv_\lambda&=&e_k \wedge v_\lambda ,\\
\psi^*_k
v_\lambda&=&\sum_{i=1}^{l(\lambda)}\frac{(-1)^{i-1}}{2}\delta_{k,\lambda_i}e_{\lambda_1}
\wedge \cdots \wedge \widehat{e}_{\lambda_i} \wedge \cdots
e_{\lambda_l}.
\end{eqnarray*}
These operators satisfy the following anti-commutation relations
$$
\begin{array}{rclccl}
\psi_i\psi^*_j+\psi^*_j\psi_i&=&\displaystyle \frac{1}{2}\delta_{i,j},&&&(i,j)\in \bbN_0^2,\\
\psi_i\psi_j+\psi_j\psi_i&=&0,&&&(i,j)\in \bbN_0^2\backslash\{(0,0)\},\\
\psi^*_i\psi^*_j+\psi^*_j\psi^*_i&=&0,&&&(i,j)\in
\bbN_0^2\backslash\{(0,0)\}.
\end{array}
$$
Observe that for any $i\in\bbN$
\begin{equation} \lb{PsiPsi*}
\psi_i \psi_i^{*}v_\lambda=
\begin{cases}
v_\lambda/2& \text {if }i \in \lambda, \\
0 & \text {otherwise}.
\end{cases}
\end{equation}

Let $\psi(z)$ be a generating function for $\psi_k$ and $\psi^*_k$:
\begin{equation}\lb{defpsi}
\psi(z)=\sum_{k \in  \bbZ} \widetilde{\psi}_k z^k,
\end{equation}
where
$$
\widetilde {\psi}_k=
\begin{cases}
\psi_k&k \geq0,\\
(-1)^k\psi^*_{-k}&k\leq-1.
\end{cases}
$$
Then the following is true
$$
\langle\widetilde{\psi}_k \widetilde{\psi}_l
v_{\emptyset},v_{\emptyset}\rangle=
\begin{cases}
{(-1)^k}/{2}& k=-l \geq1,\\
{1}/{4}& k=l=0,\\
0&{\text {otherwise,}}
\end{cases}
$$
and
\begin{eqnarray*}
\langle \psi(z) \psi(w)v_{\emptyset},v_{\emptyset} \rangle&=&
\sum_{k=0}^{\infty}\langle \psi^*_k\psi_k
v_{\emptyset},v_{\emptyset}\rangle \left( -\frac{w}{z}\right)^k\\
&=&\frac{z-w}{4(z+w)}, \; \; \;\;\; \text{for }|z|>|w|.
\end{eqnarray*}

For every odd positive integer $n$ we introduce an operator
$\alpha_n$ and its adjoint $\alpha_{-n}=\alpha^{*}_n$:
$$
\alpha_n=\sum_{k \in
\bbZ}(-1)^k\widetilde{\psi}_{k-n}\widetilde{\psi}_{-k},
$$
$$
\alpha_{-n}=\sum_{k \in
\bbZ}(-1)^k\widetilde{\psi}_{n-k}\widetilde{\psi}_{k}.
$$
Then the following commutation relations are true
$$
\left[ \alpha_n,\alpha_m\right]=\frac{n}{2}\delta_{n,-m},
$$
$$
[\alpha_n,\psi(z)]=z^n\psi(z),
$$
for all odd integers $n$ and $m$, where
$[\alpha,\beta]=\alpha\beta-\beta\alpha$.

We introduce two operators $\Gamma_{+}$ and its adjoint
$\Gamma_{-}$. They are defined by
$$
\Gamma_{\pm}(x)=\exp\left[\sum_{n=1,3,5,\dots}\frac{2p_n(x)}{n}\alpha_{\pm
n}\right].
$$
Here $p_n\in\Lambda$ are the power sums. Then one has:

\begin{equation} \lb{G+v0}
\Gamma_{+}(x) v_{\emptyset}=v_{\emptyset},
\end{equation}

\begin{equation}\lb{G+G-}
\Gamma_{+}(y) \Gamma_{-}(x)=H(x,y)\Gamma_{-}(x) \Gamma_{+}(y),
\end{equation}

\begin{equation} \lb{GFunct} \Gamma_{\pm}(x) \psi(z)=F(x;z^{\pm1})
\psi(z) \Gamma_{\pm}(x),
\end{equation}
where $H(x,y)$ and $F(x;z)$ are given with (\ref{Hfrac}) and
(\ref{Q}), respectively.

The connection between the described Fock space and skew Schur $P$
and $Q$ functions comes from the following:

\begin{equation} \lb{G-}
\Gamma_{-}(x) v_\mu=\sum_{\lambda \; {\rm strict}}Q_{\lambda /
\mu}(x)v_{\lambda},
\end{equation}

\begin{equation} \lb{G+}
\Gamma_{+}(x) v_\lambda=\sum_{\mu \; {\rm strict}}P_{\lambda /
\mu}(x)v_{\mu}.
\end{equation}

We conclude this appendix with another proof of Proposition
\ref{QPQP}.
\begin{proof}(Proposition \ref{QPQP})
\begin{align*}
\langle \Gamma_{-}(x)v_{\mu},\Gamma_{-}(y)v_ {\lambda} \rangle&=
\biggl< \sum_{\lambda \; {\rm strict}}Q_{\lambda /
\mu}(x)v_{\lambda},
\sum_{\lambda \; {\rm strict}}Q_{\lambda / \nu}(y)v_{\lambda} \biggr>\\
&=2^{-l(\nu)}\sum_{\lambda \; {\rm strict}}Q_{\lambda /
\mu}(x)P_{\lambda / \nu}(y)
\end{align*}

\begin{align*}
\langle \Gamma_{+}(y)v_{\mu},\Gamma_{+}(x)v_ {\nu} \rangle&=\biggl<
\sum_{\tau \; {\rm strict}}P_{\mu / \tau}(y)v_{\tau},
\sum_{\tau \; {\rm strict}}P_{\nu / \tau}(x)v_{\tau} \biggr>\\
&=2^{-l(\nu)}\sum_{\tau \; {\rm strict}}P_{\mu / \tau}(y)Q_{\nu /
\tau}(x)
\end{align*}

By (\ref{G+G-}) we have
$$\langle \Gamma_{+}(y) \Gamma_{-}(x)v_{\mu},v_{\nu} \rangle=
H(x,y) \langle \Gamma_{-}(x) \Gamma_{+}(y)v_{\mu}v_{\nu} \rangle$$
which implies
$$\sum_{\lambda \; {\rm strict}}Q_{\lambda / \mu}(x)P_{\lambda / \nu}(y)=H(x,y)\sum_{\tau \; {\rm strict}}Q_{\nu / \tau}(x)P_{\mu /
\tau}(y)$$
\end{proof}

\bigskip


\end{document}